\documentclass[amsmath,amssymb,10pt,aps,notitlepage,prb,twocolumn,superscriptaddress,nofootinbib,longbibliography]{revtex4-2} 
\usepackage{lipsum}
\usepackage{float}
\usepackage{multirow}
\usepackage{latexsym}
\usepackage{amsmath}
\usepackage{amsthm}
\usepackage{amssymb}
\usepackage{epstopdf} 
\usepackage{enumitem}
\usepackage{setspace}
\usepackage{dcolumn}
\usepackage{bm}
\usepackage{setspace} 
\usepackage{slashed}
\usepackage{color}
\usepackage{tikz}
\usepackage{tikz-3dplot}
\usepackage{braket}
\usepackage{cancel}
\usepackage{graphicx}
\usepackage{balance}
\usepackage{pbox}

\usepackage{placeins}
\usepackage{lipsum}
\usepackage{mathtools}
\graphicspath{{Figures/}}

\usetikzlibrary{shapes,snakes,arrows,chains,matrix,positioning,scopes,calc}
\usetikzlibrary{decorations.markings} 

\usepackage{moreverb}
\allowdisplaybreaks 
\usepackage{color}
\usepackage[colorlinks=true,hypertexnames=false]{hyperref}
\usepackage[resetlabels]{multibib}
\newcites{S}{References (Supplementary)}

\makeatletter
\AtBeginDocument{%
  \let\OM@endthebibliography\endthebibliography
  \def\endthebibliography{\bibitem@fin\OM@endthebibliography}%
}
\makeatother

\begin{document}

\let\SMsavedaddtocontents\addtocontents
\renewcommand{\addtocontents}[2]{%
  \def\SMtocfile{#1}%
  \def\SMtocname{toc}%
  \ifx\SMtocfile\SMtocname
  \else
    \SMsavedaddtocontents{#1}{#2}%
  \fi
}

\title{Thermal Phase Transition of Three Dimensional Orthogonal Metals }

\author{Inho Song}
\thanks{These authors contributed equally to this work.}
\affiliation{Department of Physics, Korea Advanced Institute of Science and Technology (KAIST), Daejeon 34141, Korea}

\author{Minsu Park}
\thanks{These authors contributed equally to this work.}
\affiliation{Department of Physics, Korea Advanced Institute of Science and Technology (KAIST), Daejeon 34141, Korea}

\author{Hanbit Oh}
\thanks{Corresponding author: \href{mailto:oh26@uwm.edu}{oh26@uwm.edu}}
\affiliation{Department of Physics, University of Wisconsin-Milwaukee, Milwaukee, Wisconsin 53201, USA}

\author{Eun-Gook Moon}
\thanks{Corresponding author: \href{mailto:egmoon@kaist.ac.kr}{egmoon@kaist.ac.kr}}
\affiliation{Department of Physics, Korea Advanced Institute of Science and Technology (KAIST), Daejeon 34141, Korea}

\date{\today}
\begin{abstract}
Orthogonal metals provide a paradigmatic example of strongly entangled metallic states whose transport and thermodynamic properties resemble those of conventional Fermi liquids, while their single-electron excitations remain fundamentally distinct. Despite considerable interest in their unusual ground-state properties, orthogonal metals in three spatial dimensions and their finite-temperature behavior remain relatively unexplored. Here, we investigate a three-dimensional orthogonal metal and characterize the thermodynamic and spectroscopic signatures associated with its finite-temperature phase transition. Starting from a simple model of fermions coupled to spin degrees of freedom, we perturbatively demonstrate that the thermal transition known to occur in the purely bosonic limit persists in the presence of coupling to a gapless Fermi surface. We then examine physical observables that change prominently between the coherent and orthogonal-metal regimes. In particular, we calculate the evolution of the physical-electron spectral function and identify characteristic signatures in angle-resolved photoemission spectroscopy. We find a pronounced contrast in low-energy spectral weight between the coherent and orthogonal-metal regimes without requiring conventional symmetry-breaking order. Our results establish finite-temperature orthogonal-metal physics as a possible mechanism for pseudogap-like behavior in three-dimensional and quasi-two-dimensional layered materials with non-negligible interlayer hopping, which may be applied to the recent experiments in $\mathrm{Sr_2 VO_3 Fe As}$.
 
\end{abstract}

\maketitle

\section{Introduction}

Understanding metallic states beyond the conventional Fermi-liquid
paradigm remains a central problem in condensed matter physics.
In a Landau Fermi liquid, low-energy excitations are long-lived
quasiparticles carrying the same quantum numbers as the microscopic
electrons, so that thermodynamic, transport, and single-particle
properties are governed by a common Fermi surface with a finite
quasiparticle residue. Strong correlations, however, can invalidate
this correspondence and produce quantum many-body states with fractionalized
excitations and emergent gauge structures~\cite{kitaev2003,kitaev2006,savary2017,zhou2017,hamma2005,
castelnovo2008}. In particular, for metallic states, the presence of fractionalized excitations and gauge structures indicate a possibility of a setting in which metallic transport can coexist with single-particle
properties qualitatively distinct from those of a conventional Fermi
liquid.

The orthogonal metal provides a particularly simple realization of
this possibility~\cite{Nandkishore2012,chen2020,hohenadler2019}.
In a slave-spin representation, the physical electron is written as
\begin{equation}
c_{i\sigma}=\tau_i^x f_{i\sigma},
\end{equation}
where $f_{i\sigma}$ carries the charge and spin of the electron and
$\tau_i^x$ is a slave-spin degree of freedom. With a local $Z_2$ gauge redundancy, the physical electron has a
finite overlap with the $f$ fermion when the slave-spin sector is in the Higgs phase, the corresponding many-body state is continuously
connected to an ordinary Fermi liquid. In the orthogonal-metal
regime, by contrast, the coherent physical-electron quasiparticle
disappears while the $f$ fermions retain a Fermi surface.
This separation leads to the defining phenomenology of an orthogonal
metal. Charge transport, compressibility, and thermodynamic
responses can remain metallic because the $f$ fermions carry charge
and possess a Fermi surface, whereas the physical-electron spectral
function can be strongly suppressed at low energies. An orthogonal metal may therefore appear pseudogapped to
single-particle spectroscopy while remaining metallic in transport.

Previous studies have mainly addressed the ground-state properties
of orthogonal metals and quantum phase transitions between
orthogonal metals and conventional Fermi liquids~\cite{Nandkishore2012,chen2020,hohenadler2019}.
Much less is understood about their finite-temperature behavior.
This question is particularly interesting in three spatial dimensions, where a deconfined $Z_2$ gauge sector can remain stable
at nonzero temperature and undergo a genuine thermal confinement
transition. 
The existence of such a finite-temperature deconfined phase and its thermal transition was partially explored in previous work by two of us using the three-dimensional toric code~\cite{oh2022}.
Yet, a central question that remains is whether this transition survives when the $Z_2$ gauge sector is coupled to a metallic Fermi surface.
In many theories of metallic criticality, gapless fermions strongly
modify order-parameter fluctuations and may change the critical
dynamics or even destabilize the transition~\cite{hertz1976,mills1993,metlitski2010qptpart2,
metlitski2010instabilities,abanov2000,abanov2003}.

In this work, we investigate the finite-temperature transition of a
three-dimensional orthogonal metal and its consequences for
single-particle spectroscopy. We consider itinerant fermions coupled
to a $Z_2$ gauge sector and first analyze the stability of the
thermal transition against the gapless Fermi surface. 
Phenomenologically, we  consider the Ising quantum critical theory where
 a dual confinement field $\phi$ is odd under the dual $Z_2$ symmetry as well-established for the pure $Z_2$ gauge theory. 
 The leading symmetry-allowed interaction is quadratic in the confinement field while a linear Yukawa-type coupling is forbidden by the dual $Z_2$ symmetry. Then, we show
perturbatively that this coupling renormalizes the parameters of the
critical theory at non-zero temperatures and shifts the transition temperature of the pure $Z_2$ gauge theory. In other words, the fermion coupling cannot  eliminate the finite-temperature transition of the pure $Z_2$ gauge theory within the perturbative
regime considered here. Furthermore, we also prove the existence of the transition temperature by considering the exactly solvable three dimensional toric code under perturbation with the linked-cluster expansion.~\cite{gelfand1990,kubo1962}.
Thus, we conclude that the three-dimensional confinement
transition remains well defined in the presence of a charge-carrying
Fermi surface at least perturbatively.

We next examine how this transition manifests itself in the
physical-electron spectrum. Our main observable is the
momentum-resolved spectral function
\begin{equation}
A_c(\bm{k},\omega)
=
-\frac{1}{\pi}
\operatorname{Im}G_c^{R}(\bm{k},\omega),
\end{equation}
which determines the angle-resolved photoemission spectroscopy (ARPES)
intensity up to the Fermi function and
photoemission matrix elements. 
In the Higgs/Fermi-liquid regime, the
slave-spin sector generates a coherent contribution to the electron
Green's function, allowing the physical electron to inherit the
Fermi surface of the $f$ fermions. Across the thermal transition,
this coherent contribution disappears. The underlying charged
fermions remain metallic, but the near-Fermi-level spectral weight
of the physical electron becomes strongly suppressed.

We calculate this spectral evolution explicitly and find a
substantial redistribution of ARPES weight across the transition between the Higgs phase and the orthogonal metal phase at low enough temperatures where the gauge flux loop excitations are significantly suppressed.
Our results suggest a route to pseudogap-like spectroscopy in
three-dimensional and quasi-layered correlated metals. In a layered
system, weak but finite interlayer coupling can stabilize a genuine
three-dimensional thermal transition while retaining strongly
anisotropic electronic behavior over a broad temperature range.
The orthogonal-metal mechanism can therefore produce a pronounced
suppression of low-energy electron spectral weight without a
conventional symmetry-breaking gap and without destroying the
charge-carrying Fermi surface.

\section{Model Hamiltonian}
\label{sec:model_hamiltonian}
\begin{figure}[tb]
    \centering    \includegraphics[width=0.99\linewidth]{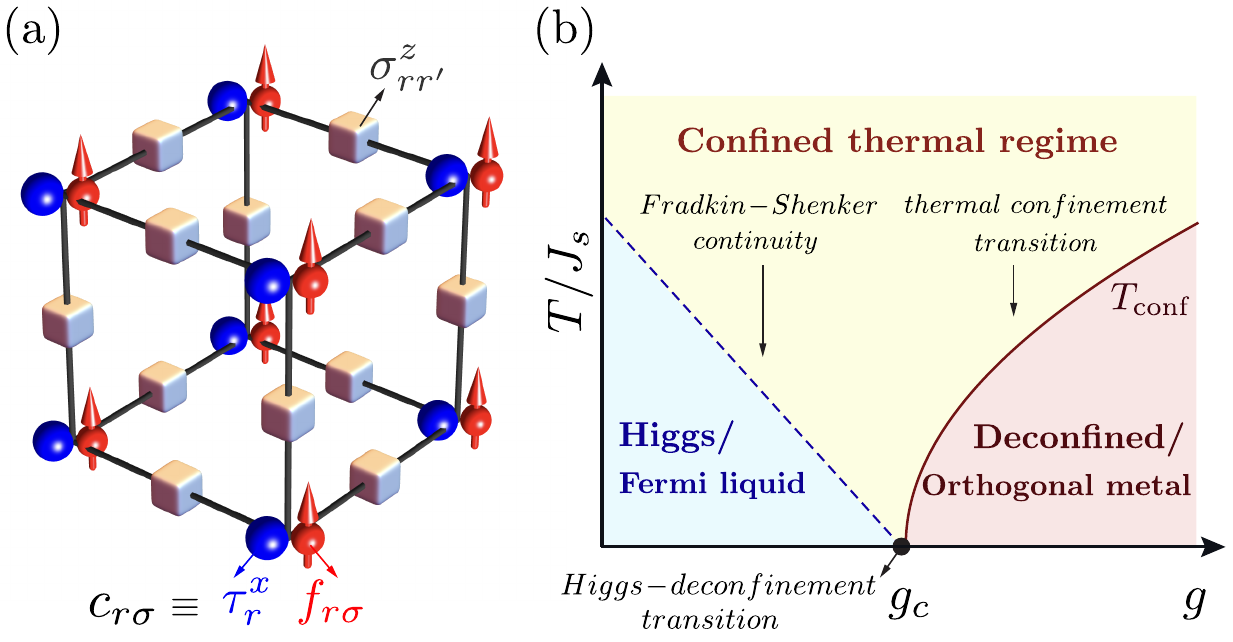}
    \caption{(a)~Schematics of three-dimensional $Z_2$ lattice model. The physical electron at site $r$ is represented as $c_{r\sigma}=\tau_r^x f_{r\sigma}$, where $\tau_r^x$ (blue sphere) is an Ising slave spin and $f_{r\sigma}$ (red sphere) is an auxiliary  fermion. The $Z_2$ gauge field $\sigma_{rr'}^z$(white cube) resides on each link. 
    (b)~Schematic finite-temperature phase diagram.
    The zero-temperature quantum critical point at $g=g_c$ separates the Higgs/Fermi-liquid and deconfined/orthogonal-metal regimes. A thermal confinement transition separates the orthogonal metal from the confined thermal regime, whereas the dashed line on the Higgs side denotes the Fradkin--Shenker continuity.
            }
    \label{fig:1}
\end{figure}
We consider a model of spinless fermions coupled to quantum spins on a three-dimensional cubic lattice. The total Hamiltonian is given by
\begin{equation}
H(g)=H_f+H_s(g),
\label{eq:H_total}
\end{equation}
which generalizes the exactly solvable orthogonal-metal model originally introduced on the square lattice~\cite{Nandkishore2012} to three spatial dimensions, as illustrated in Fig.~\ref{fig:1}(a).

The fermionic Hamiltonian is
\begin{equation}
H_f=
-\frac{t}{2}\sum_{\langle r,r'\rangle}
\left(
c_r^\dagger \sigma_{rr'}^z c_{r'}
+\mathrm{H.c.}
\right)
-\mu\sum_r c_r^\dagger c_r ,
\label{eq:H_f_original}
\end{equation}
where $c_r^\dagger$ creates a spinless fermion at site $r$, $t$ is the nearest-neighbor hopping amplitude, and $\mu$ is the chemical potential. The Pauli operators $\sigma_{rr'}^\alpha$, with $\alpha=x,y,z$, describe quantum spins residing on the links $\langle r,r'\rangle$ of the cubic lattice.

The Hamiltonian for the link spins is
\begin{equation}
H_s(g)=
-J_s\sum_{\langle r,r'\rangle}\sigma_{rr'}^z
-gJ_s\sum_r
(-1)^{c_r^\dagger c_r}
A_r
,
\label{eq:H_s_original}
\end{equation}
where $J_s>0$. Here, we define the star operator centered at site $r$ as
\begin{equation}
A_r = \prod_{r' \in \mathrm{n.n.}(r)}\sigma^x_{rr'}.
\end{equation}
When the star itself is used as the index, as in Sec.~III B,
we equivalently denote the same operator by $A_s$.

For each elementary plaquette $p$, we define the conserved $Z_2$ flux operator
\begin{equation}
B_p=
\prod_{\langle r, r' \rangle \in\partial p}
\sigma_{r r'}^z ,
\label{eq:flux_operator}
\end{equation}
where $\partial p$ denotes the set of links forming the boundary of the plaquette. Since
\begin{equation}
[B_p,H(g)]=0,
\end{equation}
the Hilbert space decomposes into sectors characterized by the eigenvalues $B_p=\pm1$. 
The total partition function is 
\begin{eqnarray}
\mathcal{Z}= {\rm Tr}_{\sigma, c}(e^{-H(g)/T}),
\end{eqnarray}
where the trace is over the entire Hamiltonian of $(\sigma^{x,y,z}_{r r^{'}}, c_r, c_r^{\dagger})$, designated by the subscript ${\sigma, c}$. In other words, all flux configurations are included in the total partition function. Note that throughout this work, we set $k_B=1$, so that $\beta=1/T$.

\subsection{Flux-free sector}
The flux-free sector is defined by the condition, 
\begin{equation}
B_p=+1
\quad
\text{for every elementary plaquette }p.
\label{eq:flux_free}
\end{equation}
Within the flux-free sector, we introduce fractionalized spin variables $\tau_r^\alpha$ residing on the sites of the cubic lattice. The link-spin operators can be represented as
\begin{align}
\sigma_{rr'}^z
=
\tau_r^x\tau_{r'}^x,
\quad 
\prod_{r'\in\mathrm{n.n.}(r)}
\sigma_{rr'}^x
=
\tau_r^z.
\label{eq:sigma_x_tau}
\end{align}
The first relation automatically satisfies the flux-free constraint, since every site variable $\tau_r^x$ appears twice in the product around an elementary plaquette:
\begin{equation}
\prod_{\langle r,r'\rangle\in\partial p}
\sigma_{rr'}^z
=
\prod_{\langle r,r'\rangle\in\partial p}
\tau_r^x\tau_{r'}^x
=1.
\end{equation}

We next define the fractionalized fermion operator
\begin{equation}
f_r=c_r\tau_r^x.
\label{eq:f_fermion}
\end{equation}
Because $(\tau_r^x)^2=1$, the inverse transformation is $c_r=f_r\tau_r^x$, and the local fermion occupation is unchanged:
\begin{equation}
c_r^\dagger c_r=f_r^\dagger f_r.
\label{eq:number_identity}
\end{equation}

We further introduce the transformed spin operators
\begin{align}
s_r^z
\equiv
\tau_r^z(-1)^{f_r^\dagger f_r},
\quad
s_r^x
\equiv
\tau_r^x.
\label{eq:tau_tilde_x}
\end{align}
These operators obey the same Pauli algebra as the original $\tau_r^\alpha$ operators. In terms of $f_r$ and $s_r^\alpha$, the Hamiltonian separates into two decoupled sectors:
\begin{align}
H_f
&=
-\frac{t}{2}\sum_{\langle r,r'\rangle}
\left(
f_r^\dagger f_{r'}
+\mathrm{H.c.}
\right)
-\mu\sum_r f_r^\dagger f_r ,
\\
H_s(g)
&=
-J_s\sum_{\langle r,r'\rangle}
s_r^xs_{r'}^x
-gJ_s\sum_rs_r^z .
\label{eq:H_s_transformed}
\end{align}
Thus, within the flux-free sector, the original interacting Hamiltonian is mapped exactly onto a system of free $f$ fermions and a transverse-field Ising model on the cubic lattice, and the partition function of the flux-free sector is 
\begin{equation}
\mathcal{Z}_{{\rm flux-free}} =
 {\rm Tr}_{f} (e^{-H_f/T}){\rm Tr}_{s} (e^{-H_s(g)/T}),
\end{equation}
where the subscripts $(s,f)$ indicates the degrees of freedom for the traces. 
Although the transformed Hamiltonian is decoupled, the physical electron remains a composite operator:
\begin{equation}
c_r=f_rs_r^x.
\label{eq:physical_electron}
\end{equation}
Consequently, physical-electron correlation functions involve contributions from both the fermionic and spin sectors.

The spin Hamiltonian in Eq.~\eqref{eq:H_s_transformed} is the transverse-field Ising model on the three-dimensional cubic lattice. The model hosts both quantum~\cite{weihong1994}  and thermal~\cite{nagai1987}. 
At zero temperature, increasing $g$ drives a quantum phase transition from an ordered phase to a disordered phase. For finite temperature, increasing temperature drives a thermal transition from the ordered phase to a high-temperature disordered phase.

The situation requires more care in two spatial dimensions. Although the corresponding two-dimensional model possesses a zero-temperature quantum phase transition, it does not exhibit a genuine finite-temperature deconfinement transition when all gauge-flux sectors are included. The apparent finite-temperature Ising transition obtained from the above mapping is an artifact of restricting the analysis to the flux-free sector. At nonzero temperature, pointlike flux excitations proliferate in two dimensions, and the flux-free sector alone does not describe the full thermal ensemble~\cite{savit1980,gregor2011,balian1975gauge2,balian1975gauge3}.

In three dimensions, by contrast, the gauge-flux excitations are extended looplike objects~\cite{savit1980,gregor2011,balian1975gauge2,balian1975gauge3}, and their proliferation can occur through a genuine finite-temperature transition~\cite{oh2022}.
The three-dimensional model therefore provides a natural setting in which to investigate the finite-temperature properties of an orthogonal metal. In the following, we focus on the physical-electron spectral function and the resulting ARPES signatures of the orthogonal-metal regime.

\section{Stability of the Thermal Transition}
\label{sec:thermal_transition}

We now discuss the finite-temperature phase structure of the three-dimensional model. The phase diagram is conveniently organized in terms of two dimensionless parameters. The first is the coupling $g$, which controls the strength of quantum fluctuations generated by the transverse-field term. The second is the reduced temperature $T/J_s$, which measures the strength of thermal fluctuations relative to the characteristic energy scale of the spin sector.

In principle, the ratio between the fermionic kinetic-energy scale and temperature, such as $t/T$, provides an additional dimensionless parameter. In the present work, however, we focus on the phase structure associated with the spin and gauge sectors. A systematic study of the effects of $t/T$, including the crossover between coherent and incoherent fermionic dynamics, is left for future work.

We already have seen within the flux-free sector, the spin Hamiltonian is mapped onto the transverse-field Ising model (See Eq.~\eqref{eq:H_s_transformed}).
At zero-temperature, this exhibits a transition from Ising-ordered phase at small $g$ to an Ising-disordered phase at large $g$. 
The physical meaning of these two phases follows from the composite representation of the electron operator.
In the Ising-ordered phase,
\begin{equation}
\left\langle s_r^x\right\rangle\neq0,
\quad Z_c\neq 0, 
\end{equation}
the physical electron has a finite overlap with the $f$ fermion. 
Here $Z_c$ denotes the quasiparticle residue of the physical electron.
The electron Green's function therefore contains a coherent quasiparticle contribution, and the resulting metallic state is continuously connected to a conventional Fermi liquid. By contrast, in the Ising-disordered phase,
\begin{equation}
\left\langle s_r^x\right\rangle=0,
\quad Z_c=0
\end{equation}
the coherent physical-electron quasiparticle weight vanishes, while the $f$ fermions retain a metallic Fermi surface. Thus, transport and thermodynamic responses associated with the $f$ fermions may remain metallic, whereas the physical-electron spectral function lacks a coherent quasiparticle pole. Provided that the $Z_2$ gauge sector remains deconfined, this phase realizes an orthogonal metal.

We now turn to the finite-temperature phase structure. At zero
temperature, the ordered--disordered distinction of the slave-spin
sector within the flux-free representation is sufficient to characterize
the Fermi-liquid and orthogonal-metal regimes. At nonzero temperature,
however, the full thermal ensemble also should include thermally excited
gauge-flux loops, so the slave-spin distinction alone no longer captures
the complete phase structure. Therefore the resulting phase diagram is more
naturally described in terms of three regimes: Higgs regime, deconfined regime, and
confined thermal regime.

For small $g$ and low $T/J_s$, the $ s^x$ spins are ordered. In the gauge-theory language, this region corresponds to the Higgs regime. Since the physical electron has a finite overlap with the $f$ fermion, the metallic state exhibits a coherent electron quasiparticle and is continuously connected to a conventional Fermi liquid.
Gauge-flux loops are dilute in this regime, and the flux-free mapping provides a useful description of the low-energy properties.

Importantly, the Higgs regime need not be separated from the high-temperature confined regime by a genuine thermodynamic phase transition. For a $Z_2$ gauge field coupled to matter in the fundamental representation, the Higgs and confined limits may be analytically connected~\cite{fradkin1979,savit1980,gregor2011,balian1975gauge2,balian1975gauge3,tupitsyn2010}. Consequently, increasing the temperature at sufficiently small $g$ can produce a smooth crossover from a low-temperature Fermi-liquid regime to a high-temperature confined and incoherent metallic regime.

For sufficiently large $g$ and low $T/J_s$, the slave spins are disordered while gauge-flux loops remain dilute. The $Z_2$ gauge field is then deconfined, and the system realizes an orthogonal metal.
The deconfined orthogonal-metal phase survives only as long as flux loops remain sufficiently dilute. Upon increasing the temperature at fixed $g>g_c$, the configurational entropy of extended flux loops eventually overcomes their energy cost. Their proliferation drives a finite-temperature transition from the deconfined orthogonal metal to a confined thermal regime.

This transition should be distinguished from a simple Ising-ordering transition. Within the low-temperature flux-free description, the orthogonal metal is already slave-spin disordered. The relevant finite-temperature transition is instead associated with the proliferation of $Z_2$ flux loops and the loss of deconfinement.

The thermal transition line may be written schematically as
\begin{equation}
T=T_\mathrm{conf}(g),
\end{equation}
where $T_\mathrm{conf}(g)$ is the confinement temperature. This line emerges from the zero-temperature quantum critical point and separates the deconfined orthogonal-metal region from the confined thermal regime. The precise shape of $T_{\mathrm{conf}}(g)$ depends on microscopic parameters and on the energetic and entropic properties of the flux loops.

The schematic structure of the phase diagram is given in Fig.~\ref{fig:1}(b). Although these three regimes are physically distinguishable, they do not necessarily represent three thermodynamically distinct phases. The Higgs and confined regimes may be smoothly connected~\cite{fradkin1979,savit1980,gregor2011,balian1975gauge2,balian1975gauge3,tupitsyn2010}, whereas the deconfined orthogonal metal is separated from them by a genuine phase boundary.

\subsection{Perturbative stability against fermionic coupling}
\label{subsec:perturbative_stability}

We now examine whether the finite-temperature confinement transition remains stable in the presence of itinerant fermions~\cite{prosko2017}. Since the transition is driven by the proliferation of $Z_2$ flux loops, the appropriate critical field is not the coarse-grained expectation value of the slave-spin operator $ s^x$. Instead, we introduce a dual Ising field $\phi$ that describes the thermal confinement transition of the gauge sector, also called Ising$^\mathrm{conf}$ transition.

In the dual description, the deconfined and confined phases are distinguished by the ordering of $\phi$. With an appropriate convention, one may write
\begin{align}
\langle \phi\rangle &=0,
&&
\text{deconfined phase},
&&
\langle \phi\rangle \neq0,
&&
\text{confined phase}, \nonumber
\end{align}
where the ordered dual phase corresponds to the proliferation of gauge-flux loops. The precise identification of the ordered side depends on the convention used for the duality transformation~\cite{wegner1971,qin2017,senthil2019,jian2019}, but the essential point is that $\phi$ represents the dual degree of freedom associated with flux-loop condensation rather than the original slave-spin order parameter.

Near the thermal confinement transition, the long-wavelength gauge sector can be represented by an effective Ising action,
\begin{equation}
S_{\phi}
= \int_{\tau, x}
\left[
\frac{1}{2}
\left(\partial_\tau\phi\right)^2
+
\frac{v_\phi^2}{2}
\left(\boldsymbol{\nabla}\phi\right)^2
+
\frac{r}{2}\phi^2
+
\frac{u}{4!}\phi^4
+\cdots
\right], \nonumber
\label{eq:dual_Ising_action}
\end{equation}
with $\int_{\tau,x} =\int_0^\beta d\tau
\int d^3x$.
Since the dual field describes non-local properties of the $Z_2$ gauge degrees of freedom, the Yukawa-type coupling to fermions is not allowed. In other words, the itinerant fermions can couple only to operators that are invariant under the dual $Z_2$ transformation, $\phi\longrightarrow-\phi$. 
The leading symmetry-allowed coupling is nothing but an energy density coupling whose form is  
\begin{equation}
S_{\mathrm{int}}
=\lambda
\int_{\tau,x}
\phi^2 \ \mathcal{O}_f , \quad 
\mathcal{O}_f= f^\dagger f. \nonumber
\label{eq:dual_fermion_coupling}
\end{equation}
Then, integrating out the fermions produces corrections to the effective action of the dual field. The quadratic coefficient is shifted according to
\begin{equation}
r
\longrightarrow
r_{\mathrm{eff}}
=
r+2\lambda
\langle\mathcal{O}_f\rangle,
\label{eq:dual_mass_shift}
\end{equation}
which changes the location of the thermal confinement transition. 
At second order in $\lambda$, the fermions generate a nonlocal interaction,
\begin{equation}
\delta S_{\phi}^{(2)}
=
-\frac{\lambda^2}{2}
\int_q
\Pi_f(q)
\left|
\left(\phi^2\right)_q
\right|^2,
\label{eq:fermion_induced_dual_interaction}
\end{equation}
where $q=(\bm{q},i\Omega_n)$ and $\Pi_f(q)$ is the correlation function of the fermionic operator $\mathcal{O}_f$,
\begin{equation}
\Pi_f(q)
=
\left\langle
\mathcal{O}_f(q)
\mathcal{O}_f(-q)
\right\rangle_{\mathrm{c}}.
\label{eq:fermion_polarization}
\end{equation}
Here the subscript $c$ denotes the connected correlation function.

At nonzero temperature, fermions have no zero-frequency Matsubara
mode, whereas the bosonic dual field possesses a static zero mode.
Consequently, the asymptotic thermal critical fluctuations are governed
by the bosonic zero mode. Provided that the static fermionic
susceptibility is regular, integrating out the fermions generates only
analytic renormalizations of the local coefficients of the classical
dual-Ising theory~\cite{chubukov2003,belitz2005}.
Then, the long-distance static
theory retains the form of a local three-dimensional Ising action,and  the resulting static effective action can be written schematically as
\begin{equation}
S_{\mathrm{eff}}^{\mathrm{cl}}
=
\int d^3x
\left[
\frac{K_{\mathrm{eff}}}{2}
\left(\boldsymbol{\nabla}\phi\right)^2
+
\frac{r_{\mathrm{eff}}}{2}\phi^2
+
\frac{u_{\mathrm{eff}}}{4!}\phi^4
+\cdots
\right].
\label{eq:classical_dual_effective_action}
\end{equation}
Perturbatively, one can safely assume that $u_{\mathrm{eff}}>0$ and that no sufficiently singular nonlocal interaction is generated. Thus, the transition remains continuous and belongs to the three-dimensional Ising universality class. The fermions therefore shift the nonuniversal parameters of the transition without introducing a new relevant operator at the Ising fixed point.

\subsection{Example of a perturbative calculation of $T_\mathrm{conf}$ }
Beyond the above phenomenological field theory approach, we present one concrete lattice model calculation of the transition temperature under a perturbation. Namely, we consider the toric code model in three spatial dimensions with a longitudinal magnetic field, whose Hamiltonian is written as 
\begin{equation}
	H_0
	=
	-J_B\sum_p B_p
	-J_A\sum_s A_s, \quad V=-h_z\sum_\ell{\sigma_{\ell}^z}.
\end{equation}
Since integrating out fermions may generate all symmetry allowed terms, we only consider the longitudinal magnetic field to be specific. Generalization to include other terms will be discussed in future works.

Introducing $K_z=\beta h_z$ and $u=\tau/\beta$, the linked-cluster
expansion~\cite{gelfand1990,kubo1962} can be written as
\begin{equation}
\ln\frac{Z}{Z_0}
=
\sum_{n=1}^{\infty}\frac{K_z^n}{n!}
\int_0^1\prod_{i=1}^n du_i\,
\left\langle
T_u O(u_1)\cdots O(u_n)
\right\rangle_{0,c},
\end{equation}
where
\begin{equation}
O(u)\equiv e^{u\beta H_0}Oe^{-u\beta H_0},
\qquad
O=\sum_\ell\sigma_\ell^z.
\end{equation}
Here, the subscript $0,c$ denotes a connected cumulant evaluated in
the unperturbed ensemble.
The corresponding change in the free-energy density $\delta f$ is simply
\begin{equation}
     \delta f=f-f_0=-\frac{T}{N}\ln\frac{Z}{Z_0}.
\end{equation}
We note that $f_0$ has the Ising-like singularity at the critical temperature, $T_{\mathrm{conf}}$, and we assume that the universality class of $f$ is the same as the one of $f_0$.

A single $\sigma_\ell^z$ anticommutes with the two star operators
adjacent to the link $\ell$. Therefore, its imaginary-time evolution
is generated entirely by the star sector, while it commutes with all
plaquette operators. Lower-order connected clusters either vanish or
produce contributions that are analytic across the flux transition.
The first contribution carrying the singularity of the flux sector
appears at fourth order, when the four links form the boundary of a
plaquette (see Eq.~\eqref{eq:flux_operator}).

After tracing over the star sector, the corresponding time-ordered
product takes the form
\begin{equation}
	\left\langle
	\mathcal{T}_{u}
	\prod_{\ell\in\partial p}
	\sigma_\ell^z(u_\ell)
	\right\rangle_{A}
	=
	C_A(\{u_l\};K_A)\,B_p,
\end{equation}
where $K_A=\beta J_A$ and $C_A^{(\pi)}$ is a dimensionless temporal kernel determined by the star sector. 
In Supplemental Material (SM), we provide the detailed form of $C_A$. 
We collect the
time-ordering and combinatorial factors into
\begin{equation}
c_4(K_A)
=
\frac{1}{4!}
\sum_{\pi\in S_4}
\int_0^1\prod_{i=1}^4du_i\,
C_A^{(\pi)}(u_1,u_2,u_3,u_4;K_A),
\label{eq:def_c4}
\end{equation}
where $\pi$ labels the assignments of the four insertions to the links
surrounding the plaquette.
Consequently, the singular part of the fourth-order free energy  can be written as 
\begin{equation}
	f_\mathrm{sing}^{(4)}
	=
	-TK_z^4\,c_4(K_A)
	\left[
	\frac{1}{N}
	\sum_p
	\langle B_p\rangle_0
	\right]_{\mathrm{sing}}.
\end{equation}

This contribution is equivalently reproduced by an effective
plaquette perturbation,
\begin{equation}
V_{\mathrm{eff}}=-J'\sum_pB_p,
\qquad
\frac{J'}{T}=c_4(K_A)K_z^4,
\end{equation}
and therefore renormalizes the dimensionless plaquette coupling up to quartic order as
\begin{equation}
K_B^{\mathrm{eff}}
=
K_B+c_4(K_A)K_z^4
\end{equation}
Let $K_{B,\mathrm{conf}}=J_B/(T_\mathrm{conf})$ denote the critical coupling of the
unperturbed model. The perturbed transition temperature $T_\mathrm{conf}'$ is
determined, to the order considered here, by
\begin{equation}
\frac{J_B}{T_\mathrm{conf}'}
+
c_4\left(\frac{J_A}{T_\mathrm{conf}'}\right)
\left(\frac{h_z}{T_\mathrm{conf}'}\right)^4
=
\frac{J_B}{T_\mathrm{conf}}
\end{equation}

\begin{figure}[b]
    \centering
\includegraphics[width=0.72\linewidth]{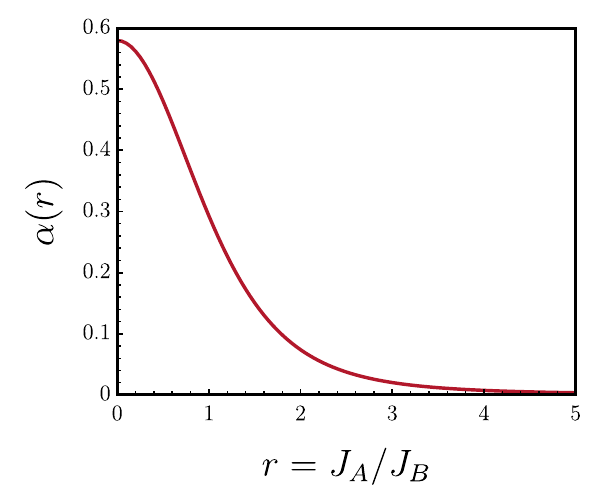}
    \vspace{-10pt}
\caption{Dimensionless coefficient $\alpha(r)$ governing the leading quartic shift of the $T_\mathrm{conf}$ in Eq.~\eqref{eq:Tconf_shift} as a function of $r=J_A/J_B$. 
We evaluate $\alpha(r)=c_4(r/\kappa)/\kappa^2$ where $c_4$ is defined in Eq.~\eqref{eq:def_c4}.
Over all range of $r$, $\alpha(r)$ remains finite and decreases monotonically with increasing $r$.
}
\label{fig:2}
\end{figure}

Expanding to the leading nonvanishing order in $h_z$, we obtain the
shift in the confinement temperature,
$\delta T_{\mathrm{conf}}=T_{\mathrm{conf}}'-T_{\mathrm{conf}}$,
as
\begin{equation}
\delta T_{\mathrm{conf}}=
\alpha(r)\frac{h_z^4}{J_B^3}
+
O\left(\frac{h_z^6}{J_B^5}\right),
\quad
\alpha(r)
=
\frac{c_4(r/\kappa)}{\kappa^2},
\label{eq:Tconf_shift}
\end{equation}
where $r\equiv J_A/J_B$ is the ratio of the star and plaquette
couplings. Here, $\kappa=1.313346(3)$ is the dimensionless critical
temperature of the unperturbed flux sector, such that
$T_{\mathrm{conf}}=\kappa J_B$ \cite{ferrenberg2018} (See SM Sec.~\ref{sec:s1} for details). The dimensionless coefficient $\alpha(r)$, which determines the magnitude of the quartic temperature shift, is
shown in Fig.~\ref{fig:2}. In the flux-only limit $J_A=0$, the operators are independent of imaginary time and $c_4(0)=1$, consequently $\alpha(0)=\kappa^{-2}$. 
Therefore longitudinal field shifts the confinement temperature only at fourth order while leaving the transition intact within the weak-field regime.

\section{Physical-electron spectral signatures}
\label{sec:spectral}
Having established the stability of the finite-temperature confinement
transition, we now examine the corresponding
single-particle signatures of electron fractionalization~\cite{Nandkishore2012,chen2020,hohenadler2019,oh2022,demedici2005,ruegg2010,hassan2010,yu2011,zhong2012,zhong2013,gazit2020}. 
Our purpose is not to describe the critical
electron spectrum near the thermal confinement transition. Instead, we
consider the Higgs/Fermi-liquid regime at $g\ll 1$ and deconfined orthogonal-metal  regime at $g\gg 1$, sufficiently far from the transition, where gauge-flux loops are dilute and the flux-free representation is applicable.
Specifically, we restrict the analysis to
\begin{equation}
T\ll\Delta_g,
\qquad
\Delta_g=\min_{\bm q}\Delta_{g,\bm q},
\label{eq:low_temperature_condition}
\end{equation}
where $\Delta_g$ denotes the minimum one-magnon gap of the slave-spin
sector.
In this regime, thermally excited slave-spin quasiparticles and gauge-flux loops
are dilute.

\subsection{Slave-spin spectral weights in the two limiting regimes}
\label{sec:slave_spin_weights}

As discussed in Sec.~\ref{sec:model_hamiltonian}, the physical electron is the composite operator
$c_r=f_rs_r^x$. Acting with $s_r^x$ on the slave-spin ground state can either leave the slave-spin sector in its ground state or create one or more slave-spin excitations. The former process offers the coherent physical-electron contribution, whereas the latter processes provides an incoherent continuum.
\\

The physical-electron spectrum is controlled by three quantities from
the longitudinal slave-spin correlator:  
\begin{itemize}
    \item Coherent weight:
    \begin{equation}
    A_0(g)=
\left|\langle0;g|s_r^x|0;g\rangle
\right|^2.    \label{eq:A0def}
\end{equation}
\item One-magnon spectral weight:
\begin{equation}
A_1(g,\bm q)
=
N\left|
\langle0;g|s_r^x|1_{\bm q};g\rangle
\right|^2.
    \label{eq:A1def}
\end{equation}
\item One-magnon excitation energy:
\begin{equation}
\Delta_{g,\bm q}
=
E_{1,\bm q}(g)-E_0(g).
    \label{eq:Deltagdef}
\end{equation}
\end{itemize}
Here $|0;g\rangle$ and $|1_{\bm q};g\rangle$ denotes respectively,
the slave-spin ground state and a normalized one-magnon state of
$H_s(g)$, with energies $E_0(g)$ and $E_{1,\bm q}(g)$. 
Here, physical meaning of $A_0$ is the coherent weight associated with leaving the
slave-spin sector in its ground state, while $A_1$ is the
momentum-resolved weight for creating a one-magnon excitation. The excitation energy
$\Delta_{g,\bm q}$ determines the energy at which the corresponding
incoherent spectral weight appears.

For $g\ll1$, the slave spins form an $x$-polarized ferromagnet~\cite{weihong1994,demedici2005,ruegg2010,hassan2010}. Acting
with $s_r^x$ therefore leaves the state nearly unchanged, and the
coherent contribution dominates. Keeping terms through second order in
$g$, we obtain:
\begin{align}
A_0(g)
&\simeq
1-\frac{g^2}{36},
\nonumber\\
A_1(g,\bm q)
&\simeq
\frac{g^2}{36},
\nonumber\\
\Delta_{g,\bm q}
&\simeq
J_s\left[
12-\frac{g^2}{12}
\left(1+\gamma_{\bm q}\right)
\right],
\label{eq:small_g_weights}
\end{align}
where $\gamma_{\bm q}
=
\cos q_x+\cos q_y+\cos q_z$.
Thus, the Higgs/Fermi-liquid regime contains a dominant coherent
component, while the one-magnon continuum carries only $O(g^2)$ and appears at an energy scale of order $12J_s$.
\\

For $g\gg1$, the slave spins instead form a $z$-polarized paramagnet~\cite{weihong1994,demedici2005,ruegg2010,hassan2010}.
In this limit, acting with $s_r^x$ necessarily flips a slave spin and
therefore creates an excitation. To leading nontrivial order in $1/g$,
\begin{align}
A_0(g)&=0,
\nonumber\\
A_1(g,\bm q)
&\simeq
1+\frac{\gamma_{\bm q}}{g},
\nonumber\\
\Delta_{g,\bm q}
&\simeq
2gJ_s-2J_s\gamma_{\bm q}.
\label{eq:large_g_weights}
\end{align}
The vanishing of $A_0$ is exact throughout the symmetry-preserving paramagnetic phase and follows directly from the global Ising symmetry.
Defining the Ising-parity operator, $P=\prod_r s_r^z$, we have
$[P,H_s(g)]=0$ and $P s_r^x P^{-1}=-s_r^x$. 
Since the paramagnetic ground state $|0;g\rangle$ is an eigenstate of
$P$, it follows that $\langle0;g|s_r^x|0;g\rangle=0$.
Therefore the orthogonal-metal regime has no coherent physical-electron contribution. Instead, its spectral weight is carried by processes involving slave-spin excitations with characteristic energy $\Delta_g$.
The full expressions through second order in $1/g$ are provided in the
SM Sec.~\ref{sec:s2} with detailed derivation.
\\

The qualitative distinction between the two regimes is therefore already visible at the level of the slave-spin spectral weights: the Higgs/Fermi-liquid regime contains a dominant coherent component, whereas the orthogonal-metal regime has no coherent contribution and places the spectral weight in the slave-spin excitation sector.

\begin{figure}[t]
\centering
\includegraphics[width=.9\linewidth]{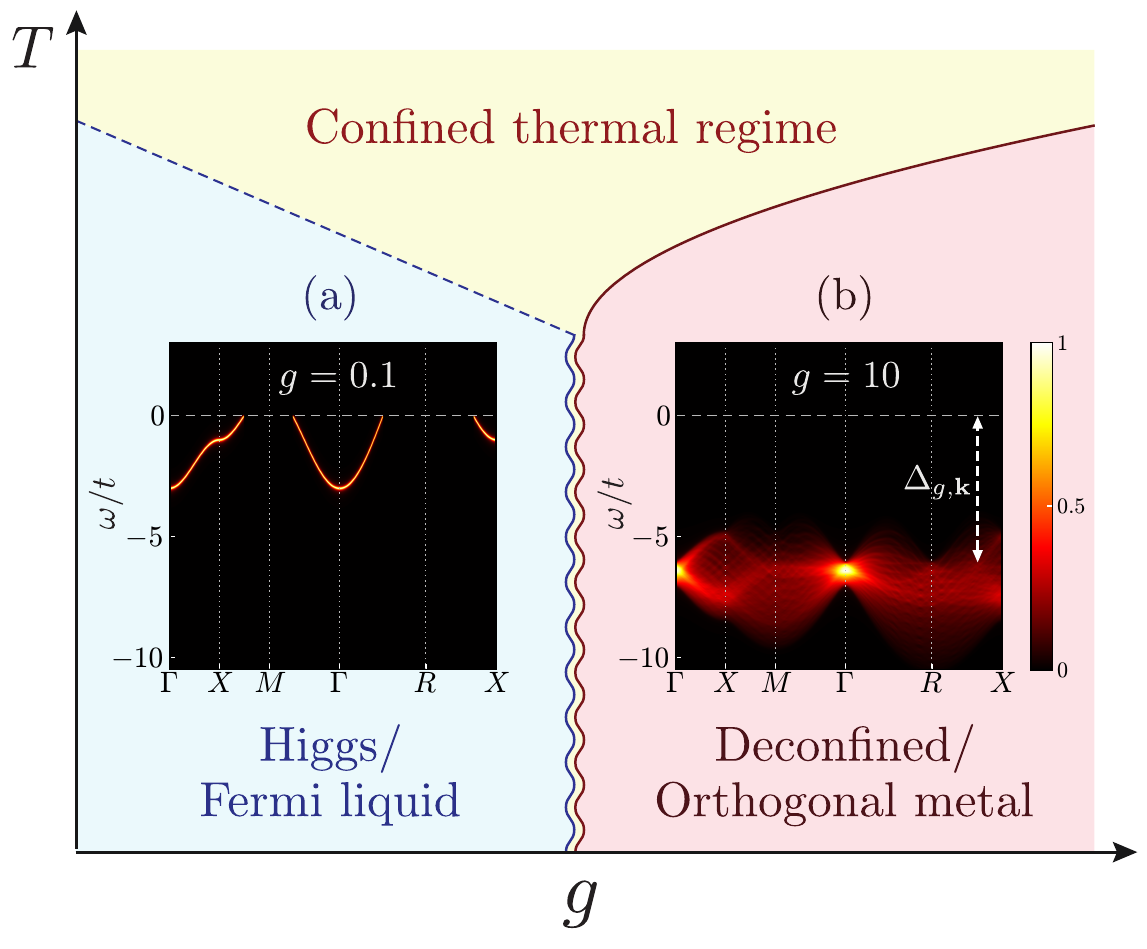}

\caption{Schematic phase diagram and representative low-temperature ARPES
spectra. The spectra are calculated within the flux-free sector at
fixed temperature $T/t=0.02$. (a) For $g=0.1$, in the Higgs/Fermi-liquid
regime, the spectrum is dominated by a sharp coherent quasiparticle
band. (b) For $g=10$, in the deconfined orthogonal-metal regime, the
coherent contribution vanishes and the spectral weight instead forms
a broad incoherent continuum at finite binding energy. The intensity
is plotted along the high-symmetry path
$\Gamma$--$X$--$M$--$\Gamma$--$R$--$X$. Each panel is normalized
separately, so their absolute intensities should not be compared.
A Lorentzian broadening $\Gamma=0.05t$ is used. We take $t>0$ as the
energy unit and set $\mu=0$ and $J_s/t=0.3$.}
\label{fig:3}
\end{figure}

\subsection{Physical-electron spectral function and ARPES}
\label{sec:electron_spectrum}
We now combine the slave-spin spectral weights with the free
$f$-fermion propagator. Since the fermionic and slave-spin sectors are
exactly decoupled in the flux-free representation, the physical-electron spectral function is obtained by convolving the free-$f$-fermion and
slave-spin spectra as in previous works~\cite{Nandkishore2012,ruegg2010,zhong2013}.
We take the nearest-neighbor f-fermion dispersion
\begin{equation}
\epsilon_{\bm k}
=
-t(\cos k_x+\cos k_y+\cos k_z)-\mu.
\end{equation}

Using the standard way of calculating the
physical-electron spectrum~\cite{zhong2013}, and inserting the
ground-state and one-magnon contributions defined above, we obtain,
for $T\ll\Delta_g$,
\begin{align}
&A_c(\bm{k},\omega;g)
=A_0(g)\delta(\omega-\epsilon_{\bm{k}})
\nonumber\\
&+\int_{\bm q}
A_1(g,\bm{k}-\bm q)
\left[1-n_F(\epsilon_{\bm q})\right]
\delta\!\left[
\omega-\epsilon_{\bm q}
-\Delta_{g,\bm{k}-\bm q}
\right]
\nonumber\\
&+\int_{\bm q}
A_1(g,\bm{k}-\bm q)
n_F(\epsilon_{\bm q})
\delta\!\left[
\omega-\epsilon_{\bm q}
+\Delta_{g,\bm{k}-\bm q}
\right].
\label{eq:physical_spectral_function}
\end{align}
where
$\int_{\bm{q}}=\int_{\mathrm{BZ}}d^3q/(2\pi)^3$ 
and $n_F(E)$ is the Fermi-Dirac distribution. Note that corrections from bosonic occupation of slave-spin states are exponentially suppressed as
$e^{-\Delta_g/T}$ under the condition, Eq.~\eqref{eq:low_temperature_condition}.
The detailed derivations are provided in SM.

The first term in Eq.~\eqref{eq:physical_spectral_function} is the coherent quasiparticle contribution inherited directly from the $f$-fermion band.
The second and third terms describe particle- and hole-like
fermionic excitations, respectively, convoluted with a
slave-spin excitation, with their thermal occupations encoded
by the Fermi--Dirac factors.
In the Higgs/Fermi-liquid regime, $A_0\simeq1$, and the physical-electron
spectrum is dominated by a sharp band that crosses the Fermi level. The
one-magnon continuum carries only a small fraction of the total weight.
In the orthogonal-metal regime, by contrast, $A_0=0$ exactly. Although
the $f$ fermions retain a metallic Fermi surface, the physical-electron
quasiparticle pole is absent, and the spectral weight is transferred to
a finite-energy continuum.

Within the sudden approximation and neglecting
momentum-dependent photoemission matrix elements, the ARPES intensity
is proportional to the occupied part of the physical-electron
spectral function~\cite{damascelli2003,sobota2021},
\begin{equation}
    I_{\rm ARPES}(\bm k,\omega;T)
    \propto
    n_F(\omega)
    A_c(\bm k,\omega).
    \label{eq:ARPES}
\end{equation}
For Fig.~\ref{fig:3}, finite energy resolution is modeled by replacing each delta function in
Eq.~\eqref{eq:physical_spectral_function} by a Lorentzian. We specified the energy resolution $\Gamma$ in the caption.

Fig.~\ref{fig:3} illustrates the spectral reconstruction. 
In the Higgs/Fermi-liquid regime, the finite
coherent weight produces a sharp band crossing the Fermi level. In
the orthogonal-metal regime, this coherent feature is absent and the
spectral weight is transferred to a broad continuum at finite binding
energy. The resulting suppression of the near-Fermi-level
photoemission intensity is the principal signature of the orthogonal-metal regime considered here.
We stress that the calculation applies the low-temperature flux-free regime and do
not include the critical broadening caused by thermally proliferating
gauge-flux loops near the confinement transition.
A complementary perturbative analysis of the physical-electron
self-energy generated by closed temporal star fluctuations in the
distinct hierarchy $E_c\gg T\gg J_s,h$ is presented in the SM Sec.~\ref{sec:s3}.

\section{Discussion and Conclusion}

The central theoretical result of this work is that the
finite-temperature confinement transition of a three-dimensional
orthogonal metal remains stable against weak coupling to a metallic
Fermi surface. Within the perturbative regime considered here, the
fermions renormalize the parameters of the effective theory and shift
the transition temperature, but do not eliminate the transition or
change its three-dimensional Ising criticality
~\cite{hertz1976,mills1993,chubukov2003,belitz2005}.

The microscopic calculations of Sec.~IV establish the corresponding
spectroscopic consequences. Since the physical electron is a
composite object, fractionalization can strongly affect
single-particle probes even when the charge-carrying $f$ fermions
remain metallic. Across the confinement transition, the coherent
physical-electron contribution is strongly suppressed, producing a
pronounced reconstruction of the momentum-resolved spectral function
and ARPES intensity
~\cite{Nandkishore2012,gazit2020}. This separation between metallic
transport and strongly modified single-particle spectroscopy is a
characteristic signature of the orthogonal-metal regime.

Although our model is not intended as a microscopic description of a
specific material, this mechanism may provide a useful framework for
pseudogap-like spectroscopic behavior coexisting with metallic
transport. One potentially relevant system is
$\rm Sr_2VO_3FeAs$, where experiments have reported electronic
reconstruction near 150K while the system remains metallic
~\cite{ok2017,lee2019,choi2017,kim2021,yu2011,qian2011}.
These observations are commonly discussed in terms of orbital or
charge ordering; finite-temperature confinement in a fractionalized
metal provides a distinct phenomenological possibility that remains
to be examined microscopically.

Several questions remain open. Our spectral calculations employ the
exactly solvable flux-free representation and therefore neglect
dynamical gauge-flux fluctuations near the critical regime. A more
complete treatment should incorporate thermally proliferating flux
loops directly into the electron Green's function. It would also be
interesting to investigate transport near the transition and the
effects of stronger fermion-gauge coupling, where the perturbative
description may eventually break down.

In conclusion, we have established a minimal framework connecting
finite-temperature gauge confinement, electron fractionalization,
and single-particle spectroscopy in a three-dimensional orthogonal
metal at non-zero temperatures. Our results demonstrate
how finite-temperature fractionalization can produce strongly
suppressed low-energy spectral weight without destroying the
underlying metallic state.

\section{Acknowledgments} 
This work was supported by the Korea Research Institute of Standards and Science (KRISS) under Grant No. KRISS-2026-GP2026-0015 and by the National Research Foundation of Korea (NRF), funded by the Korean government (Ministry of Science and ICT, MSIT), under Grant Nos. RS-2022-NR068223, RS-2024-00451261, RS-2025-00559286, RS-2023-00281839, RS-2023-00256050, and RS-2025-08542968. 
H.O. is supported by the U.S. Department of Energy, Office of Basic Energy Sciences, Division of Materials Sciences and Engineering under Award No. DE-SC0021971.

\section{Data Availability}
The data that support the findings of this article are available from the authors upon reasonable request.

\makeatletter
\let\NAT@bibsetnum\NATx@bibsetnum
\let\@bibsetup\NATx@bibsetnum
\setbiblabelwidth{99}
\makeatother
\bibliographystyle{apsrev4-2}
\bibliography{refs}

\begin{thebibliography}{8}%
\makeatletter
\providecommand \@ifxundefined [1]{%
 \@ifx{#1\undefined}
}%
\providecommand \@ifnum [1]{%
 \ifnum #1\expandafter \@firstoftwo
 \else \expandafter \@secondoftwo
 \fi
}%
\providecommand \@ifx [1]{%
 \ifx #1\expandafter \@firstoftwo
 \else \expandafter \@secondoftwo
 \fi
}%
\providecommand \natexlab [1]{#1}%
\providecommand \enquote  [1]{``#1''}%
\providecommand \bibnamefont  [1]{#1}%
\providecommand \bibfnamefont [1]{#1}%
\providecommand \citenamefont [1]{#1}%
\providecommand \href@noop [0]{\@secondoftwo}%
\providecommand \href [0]{\begingroup \@sanitize@url \@href}%
\providecommand \@href[1]{\@@startlink{#1}\@@href}%
\providecommand \@@href[1]{\endgroup#1\@@endlink}%
\providecommand \@sanitize@url [0]{\catcode `\\12\catcode `\$12\catcode
  `\&12\catcode `\#12\catcode `\^12\catcode `\_12\catcode `\%12\relax}%
\providecommand \@@startlink[1]{}%
\providecommand \@@endlink[0]{}%
\providecommand \url  [0]{\begingroup\@sanitize@url \@url }%
\providecommand \@url [1]{\endgroup\@href {#1}{\urlprefix }}%
\providecommand \urlprefix  [0]{URL }%
\providecommand \Eprint [0]{\href }%
\providecommand \doibase [0]{https://doi.org/}%
\providecommand \selectlanguage [0]{\@gobble}%
\providecommand \bibinfo  [0]{\@secondoftwo}%
\providecommand \bibfield  [0]{\@secondoftwo}%
\providecommand \translation [1]{[#1]}%
\providecommand \BibitemOpen [0]{}%
\providecommand \bibitemStop [0]{}%
\providecommand \bibitemNoStop [0]{.\EOS\space}%
\providecommand \EOS [0]{\spacefactor3000\relax}%
\providecommand \BibitemShut  [1]{\csname bibitem#1\endcsname}%
\let\auto@bib@innerbib\@empty
\bibitem [{\citenamefont {Kubo}(1962)}]{S_kubo1962}%
  \BibitemOpen
  \bibfield  {author} {\bibinfo {author} {\bibfnamefont {R.}~\bibnamefont
  {Kubo}},\ }\href {https://doi.org/10.1143/JPSJ.17.1100} {\bibfield  {journal}
  {\bibinfo  {journal} {J. Phys. Soc. Jpn.}\ }\textbf {\bibinfo {volume}
  {17}},\ \bibinfo {pages} {1100} (\bibinfo {year} {1962})}\BibitemShut
  {NoStop}%
\bibitem [{\citenamefont {Gelfand}\ \emph {et~al.}(1990)\citenamefont
  {Gelfand}, \citenamefont {Singh},\ and\ \citenamefont
  {Huse}}]{S_gelfand1990}%
  \BibitemOpen
  \bibfield  {author} {\bibinfo {author} {\bibfnamefont {M.~P.}\ \bibnamefont
  {Gelfand}}, \bibinfo {author} {\bibfnamefont {R.~R.~P.}\ \bibnamefont
  {Singh}},\ and\ \bibinfo {author} {\bibfnamefont {D.~A.}\ \bibnamefont
  {Huse}},\ }\href {https://doi.org/10.1007/BF01334744} {\bibfield  {journal}
  {\bibinfo  {journal} {J. Stat. Phys.}\ }\textbf {\bibinfo {volume} {59}},\
  \bibinfo {pages} {1093} (\bibinfo {year} {1990})}\BibitemShut {NoStop}%
\bibitem [{\citenamefont {Wegner}(1971)}]{S_wegner1971}%
  \BibitemOpen
  \bibfield  {author} {\bibinfo {author} {\bibfnamefont {F.~J.}\ \bibnamefont
  {Wegner}},\ }\href {https://doi.org/10.1063/1.1665530} {\bibfield  {journal}
  {\bibinfo  {journal} {Journal of Mathematical Physics}\ }\textbf {\bibinfo
  {volume} {12}},\ \bibinfo {pages} {2259} (\bibinfo {year}
  {1971})}\BibitemShut {NoStop}%
\bibitem [{\citenamefont {Ferrenberg}\ \emph {et~al.}(2018)\citenamefont
  {Ferrenberg}, \citenamefont {Xu},\ and\ \citenamefont
  {Landau}}]{S_ferrenberg2018}%
  \BibitemOpen
  \bibfield  {author} {\bibinfo {author} {\bibfnamefont {A.~M.}\ \bibnamefont
  {Ferrenberg}}, \bibinfo {author} {\bibfnamefont {J.}~\bibnamefont {Xu}},\
  and\ \bibinfo {author} {\bibfnamefont {D.~P.}\ \bibnamefont {Landau}},\
  }\href {https://doi.org/10.1103/PhysRevE.97.043301} {\bibfield  {journal}
  {\bibinfo  {journal} {Phys. Rev. E}\ }\textbf {\bibinfo {volume} {97}},\
  \bibinfo {pages} {043301} (\bibinfo {year} {2018})}\BibitemShut {NoStop}%
\bibitem [{\citenamefont {{Zheng Weihong}}\ \emph {et~al.}(1994)\citenamefont
  {{Zheng Weihong}}, \citenamefont {Oitmaa},\ and\ \citenamefont
  {Hamer}}]{S_weihong1994}%
  \BibitemOpen
  \bibfield  {author} {\bibinfo {author} {\bibnamefont {{Zheng Weihong}}},
  \bibinfo {author} {\bibfnamefont {J.}~\bibnamefont {Oitmaa}},\ and\ \bibinfo
  {author} {\bibfnamefont {C.~J.}\ \bibnamefont {Hamer}},\ }\href
  {https://doi.org/10.1088/0305-4470/27/16/010} {\bibfield  {journal} {\bibinfo
   {journal} {J. Phys. A: Math. Gen.}\ }\textbf {\bibinfo {volume} {27}},\
  \bibinfo {pages} {5425} (\bibinfo {year} {1994})}\BibitemShut {NoStop}%
\bibitem [{\citenamefont {Bravyi}\ \emph {et~al.}(2011)\citenamefont {Bravyi},
  \citenamefont {DiVincenzo},\ and\ \citenamefont {Loss}}]{S_bravyi2011}%
  \BibitemOpen
  \bibfield  {author} {\bibinfo {author} {\bibfnamefont {S.}~\bibnamefont
  {Bravyi}}, \bibinfo {author} {\bibfnamefont {D.~P.}\ \bibnamefont
  {DiVincenzo}},\ and\ \bibinfo {author} {\bibfnamefont {D.}~\bibnamefont
  {Loss}},\ }\href {https://doi.org/10.1016/j.aop.2011.06.004} {\bibfield
  {journal} {\bibinfo  {journal} {Ann. Phys.}\ }\textbf {\bibinfo {volume}
  {326}},\ \bibinfo {pages} {2793} (\bibinfo {year} {2011})}\BibitemShut
  {NoStop}%
\bibitem [{\citenamefont {Nandkishore}\ \emph {et~al.}(2012)\citenamefont
  {Nandkishore}, \citenamefont {Metlitski},\ and\ \citenamefont
  {Senthil}}]{S_Nandkishore2012}%
  \BibitemOpen
  \bibfield  {author} {\bibinfo {author} {\bibfnamefont {R.}~\bibnamefont
  {Nandkishore}}, \bibinfo {author} {\bibfnamefont {M.~A.}\ \bibnamefont
  {Metlitski}},\ and\ \bibinfo {author} {\bibfnamefont {T.}~\bibnamefont
  {Senthil}},\ }\href {https://doi.org/10.1103/PhysRevB.86.045128} {\bibfield
  {journal} {\bibinfo  {journal} {Phys. Rev. B}\ }\textbf {\bibinfo {volume}
  {86}},\ \bibinfo {pages} {045128} (\bibinfo {year} {2012})}\BibitemShut
  {NoStop}%
\bibitem [{\citenamefont {R{\"u}egg}\ \emph {et~al.}(2010)\citenamefont
  {R{\"u}egg}, \citenamefont {Huber},\ and\ \citenamefont
  {Sigrist}}]{S_ruegg2010}%
  \BibitemOpen
  \bibfield  {author} {\bibinfo {author} {\bibfnamefont {A.}~\bibnamefont
  {R{\"u}egg}}, \bibinfo {author} {\bibfnamefont {S.~D.}\ \bibnamefont
  {Huber}},\ and\ \bibinfo {author} {\bibfnamefont {M.}~\bibnamefont
  {Sigrist}},\ }\href {https://doi.org/10.1103/PhysRevB.81.155118} {\bibfield
  {journal} {\bibinfo  {journal} {Phys. Rev. B}\ }\textbf {\bibinfo {volume}
  {81}},\ \bibinfo {pages} {155118} (\bibinfo {year} {2010})}\BibitemShut
  {NoStop}%
\end{thebibliography}%


\begin{thebibliography}{48}%
\makeatletter
\providecommand \@ifxundefined [1]{%
 \@ifx{#1\undefined}
}%
\providecommand \@ifnum [1]{%
 \ifnum #1\expandafter \@firstoftwo
 \else \expandafter \@secondoftwo
 \fi
}%
\providecommand \@ifx [1]{%
 \ifx #1\expandafter \@firstoftwo
 \else \expandafter \@secondoftwo
 \fi
}%
\providecommand \natexlab [1]{#1}%
\providecommand \enquote  [1]{``#1''}%
\providecommand \bibnamefont  [1]{#1}%
\providecommand \bibfnamefont [1]{#1}%
\providecommand \citenamefont [1]{#1}%
\providecommand \href@noop [0]{\@secondoftwo}%
\providecommand \href [0]{\begingroup \@sanitize@url \@href}%
\providecommand \@href[1]{\@@startlink{#1}\@@href}%
\providecommand \@@href[1]{\endgroup#1\@@endlink}%
\providecommand \@sanitize@url [0]{\catcode `\\12\catcode `\$12\catcode
  `\&12\catcode `\#12\catcode `\^12\catcode `\_12\catcode `\%12\relax}%
\providecommand \@@startlink[1]{}%
\providecommand \@@endlink[0]{}%
\providecommand \url  [0]{\begingroup\@sanitize@url \@url }%
\providecommand \@url [1]{\endgroup\@href {#1}{\urlprefix }}%
\providecommand \urlprefix  [0]{URL }%
\providecommand \Eprint [0]{\href }%
\providecommand \doibase [0]{https://doi.org/}%
\providecommand \selectlanguage [0]{\@gobble}%
\providecommand \bibinfo  [0]{\@secondoftwo}%
\providecommand \bibfield  [0]{\@secondoftwo}%
\providecommand \translation [1]{[#1]}%
\providecommand \BibitemOpen [0]{}%
\providecommand \bibitemStop [0]{}%
\providecommand \bibitemNoStop [0]{.\EOS\space}%
\providecommand \EOS [0]{\spacefactor3000\relax}%
\providecommand \BibitemShut  [1]{\csname bibitem#1\endcsname}%
\let\auto@bib@innerbib\@empty
\bibitem [{\citenamefont {Kitaev}(2003)}]{kitaev2003}%
  \BibitemOpen
  \bibfield  {author} {\bibinfo {author} {\bibfnamefont {A.~Y.}\ \bibnamefont
  {Kitaev}},\ }\href {https://doi.org/10.1016/S0003-4916(02)00018-0} {\bibfield
   {journal} {\bibinfo  {journal} {Annals of physics}\ }\textbf {\bibinfo
  {volume} {303}},\ \bibinfo {pages} {2} (\bibinfo {year} {2003})}\BibitemShut
  {NoStop}%
\bibitem [{\citenamefont {Kitaev}(2006)}]{kitaev2006}%
  \BibitemOpen
  \bibfield  {author} {\bibinfo {author} {\bibfnamefont {A.}~\bibnamefont
  {Kitaev}},\ }\href {https://doi.org/10.1016/j.aop.2005.10.005} {\bibfield
  {journal} {\bibinfo  {journal} {Annals of Physics}\ }\textbf {\bibinfo
  {volume} {321}},\ \bibinfo {pages} {2} (\bibinfo {year} {2006})}\BibitemShut
  {NoStop}%
\bibitem [{\citenamefont {Savary}\ and\ \citenamefont
  {Balents}(2017)}]{savary2017}%
  \BibitemOpen
  \bibfield  {author} {\bibinfo {author} {\bibfnamefont {L.}~\bibnamefont
  {Savary}}\ and\ \bibinfo {author} {\bibfnamefont {L.}~\bibnamefont
  {Balents}},\ }\href {https://doi.org/10.1088/0034-4885/80/1/016502}
  {\bibfield  {journal} {\bibinfo  {journal} {Reports on Progress in Physics}\
  }\textbf {\bibinfo {volume} {80}},\ \bibinfo {pages} {016502} (\bibinfo
  {year} {2017})}\BibitemShut {NoStop}%
\bibitem [{\citenamefont {Zhou}\ \emph {et~al.}(2017)\citenamefont {Zhou},
  \citenamefont {Kanoda},\ and\ \citenamefont {Ng}}]{zhou2017}%
  \BibitemOpen
  \bibfield  {author} {\bibinfo {author} {\bibfnamefont {Y.}~\bibnamefont
  {Zhou}}, \bibinfo {author} {\bibfnamefont {K.}~\bibnamefont {Kanoda}},\ and\
  \bibinfo {author} {\bibfnamefont {T.-K.}\ \bibnamefont {Ng}},\ }\href
  {https://doi.org/10.1103/RevModPhys.89.025003} {\bibfield  {journal}
  {\bibinfo  {journal} {Rev. Mod. Phys.}\ }\textbf {\bibinfo {volume} {89}},\
  \bibinfo {pages} {025003} (\bibinfo {year} {2017})}\BibitemShut {NoStop}%
\bibitem [{\citenamefont {Hamma}\ \emph {et~al.}(2005)\citenamefont {Hamma},
  \citenamefont {Zanardi},\ and\ \citenamefont {Wen}}]{hamma2005}%
  \BibitemOpen
  \bibfield  {author} {\bibinfo {author} {\bibfnamefont {A.}~\bibnamefont
  {Hamma}}, \bibinfo {author} {\bibfnamefont {P.}~\bibnamefont {Zanardi}},\
  and\ \bibinfo {author} {\bibfnamefont {X.-G.}\ \bibnamefont {Wen}},\ }\href
  {https://doi.org/10.1103/PhysRevB.72.035307} {\bibfield  {journal} {\bibinfo
  {journal} {Phys. Rev. B}\ }\textbf {\bibinfo {volume} {72}},\ \bibinfo
  {pages} {035307} (\bibinfo {year} {2005})}\BibitemShut {NoStop}%
\bibitem [{\citenamefont {Castelnovo}\ and\ \citenamefont
  {Chamon}(2008)}]{castelnovo2008}%
  \BibitemOpen
  \bibfield  {author} {\bibinfo {author} {\bibfnamefont {C.}~\bibnamefont
  {Castelnovo}}\ and\ \bibinfo {author} {\bibfnamefont {C.}~\bibnamefont
  {Chamon}},\ }\href {https://doi.org/10.1103/PhysRevB.78.155120} {\bibfield
  {journal} {\bibinfo  {journal} {Phys. Rev. B}\ }\textbf {\bibinfo {volume}
  {78}},\ \bibinfo {pages} {155120} (\bibinfo {year} {2008})}\BibitemShut
  {NoStop}%
\bibitem [{\citenamefont {Nandkishore}\ \emph {et~al.}(2012)\citenamefont
  {Nandkishore}, \citenamefont {Metlitski},\ and\ \citenamefont
  {Senthil}}]{Nandkishore2012}%
  \BibitemOpen
  \bibfield  {author} {\bibinfo {author} {\bibfnamefont {R.}~\bibnamefont
  {Nandkishore}}, \bibinfo {author} {\bibfnamefont {M.~A.}\ \bibnamefont
  {Metlitski}},\ and\ \bibinfo {author} {\bibfnamefont {T.}~\bibnamefont
  {Senthil}},\ }\href {https://doi.org/10.1103/PhysRevB.86.045128} {\bibfield
  {journal} {\bibinfo  {journal} {Phys. Rev. B}\ }\textbf {\bibinfo {volume}
  {86}},\ \bibinfo {pages} {045128} (\bibinfo {year} {2012})}\BibitemShut
  {NoStop}%
\bibitem [{\citenamefont {Chen}\ \emph {et~al.}(2020)\citenamefont {Chen},
  \citenamefont {Xu}, \citenamefont {Qi},\ and\ \citenamefont
  {Meng}}]{chen2020}%
  \BibitemOpen
  \bibfield  {author} {\bibinfo {author} {\bibfnamefont {C.}~\bibnamefont
  {Chen}}, \bibinfo {author} {\bibfnamefont {X.~Y.}\ \bibnamefont {Xu}},
  \bibinfo {author} {\bibfnamefont {Y.}~\bibnamefont {Qi}},\ and\ \bibinfo
  {author} {\bibfnamefont {Z.~Y.}\ \bibnamefont {Meng}},\ }\href
  {https://doi.org/10.1088/0256-307X/37/4/047103} {\bibfield  {journal}
  {\bibinfo  {journal} {Chinese Physics Letters}\ }\textbf {\bibinfo {volume}
  {37}},\ \bibinfo {pages} {047103} (\bibinfo {year} {2020})}\BibitemShut
  {NoStop}%
\bibitem [{\citenamefont {Hohenadler}\ and\ \citenamefont
  {Assaad}(2019)}]{hohenadler2019}%
  \BibitemOpen
  \bibfield  {author} {\bibinfo {author} {\bibfnamefont {M.}~\bibnamefont
  {Hohenadler}}\ and\ \bibinfo {author} {\bibfnamefont {F.~F.}\ \bibnamefont
  {Assaad}},\ }\href {https://doi.org/10.1103/PhysRevB.100.125133} {\bibfield
  {journal} {\bibinfo  {journal} {Phys. Rev. B}\ }\textbf {\bibinfo {volume}
  {100}},\ \bibinfo {pages} {125133} (\bibinfo {year} {2019})}\BibitemShut
  {NoStop}%
\bibitem [{\citenamefont {Oh}\ and\ \citenamefont {Moon}(2022)}]{oh2022}%
  \BibitemOpen
  \bibfield  {author} {\bibinfo {author} {\bibfnamefont {H.}~\bibnamefont
  {Oh}}\ and\ \bibinfo {author} {\bibfnamefont {E.-G.}\ \bibnamefont {Moon}},\
  }\href {https://doi.org/10.1103/PhysRevB.106.L140504} {\bibfield  {journal}
  {\bibinfo  {journal} {Phys. Rev. B}\ }\textbf {\bibinfo {volume} {106}},\
  \bibinfo {pages} {L140504} (\bibinfo {year} {2022})}\BibitemShut {NoStop}%
\bibitem [{\citenamefont {Hertz}(1976)}]{hertz1976}%
  \BibitemOpen
  \bibfield  {author} {\bibinfo {author} {\bibfnamefont {J.~A.}\ \bibnamefont
  {Hertz}},\ }\href {https://doi.org/10.1103/PhysRevB.14.1165} {\bibfield
  {journal} {\bibinfo  {journal} {Phys. Rev. B}\ }\textbf {\bibinfo {volume}
  {14}},\ \bibinfo {pages} {1165} (\bibinfo {year} {1976})}\BibitemShut
  {NoStop}%
\bibitem [{\citenamefont {Millis}(1993)}]{mills1993}%
  \BibitemOpen
  \bibfield  {author} {\bibinfo {author} {\bibfnamefont {A.~J.}\ \bibnamefont
  {Millis}},\ }\href {https://doi.org/10.1103/PhysRevB.48.7183} {\bibfield
  {journal} {\bibinfo  {journal} {Phys. Rev. B}\ }\textbf {\bibinfo {volume}
  {48}},\ \bibinfo {pages} {7183} (\bibinfo {year} {1993})}\BibitemShut
  {NoStop}%
\bibitem [{\citenamefont {Metlitski}\ and\ \citenamefont
  {Sachdev}(2010{\natexlab{a}})}]{metlitski2010qptpart2}%
  \BibitemOpen
  \bibfield  {author} {\bibinfo {author} {\bibfnamefont {M.~A.}\ \bibnamefont
  {Metlitski}}\ and\ \bibinfo {author} {\bibfnamefont {S.}~\bibnamefont
  {Sachdev}},\ }\href {https://doi.org/10.1103/PhysRevB.82.075128} {\bibfield
  {journal} {\bibinfo  {journal} {Phys. Rev. B}\ }\textbf {\bibinfo {volume}
  {82}},\ \bibinfo {pages} {075128} (\bibinfo {year}
  {2010}{\natexlab{a}})}\BibitemShut {NoStop}%
\bibitem [{\citenamefont {Metlitski}\ and\ \citenamefont
  {Sachdev}(2010{\natexlab{b}})}]{metlitski2010instabilities}%
  \BibitemOpen
  \bibfield  {author} {\bibinfo {author} {\bibfnamefont {M.~A.}\ \bibnamefont
  {Metlitski}}\ and\ \bibinfo {author} {\bibfnamefont {S.}~\bibnamefont
  {Sachdev}},\ }\href {https://doi.org/10.1088/1367-2630/12/10/105007}
  {\bibfield  {journal} {\bibinfo  {journal} {New Journal of Physics}\ }\textbf
  {\bibinfo {volume} {12}},\ \bibinfo {pages} {105007} (\bibinfo {year}
  {2010}{\natexlab{b}})}\BibitemShut {NoStop}%
\bibitem [{\citenamefont {Abanov}\ and\ \citenamefont
  {Chubukov}(2000)}]{abanov2000}%
  \BibitemOpen
  \bibfield  {author} {\bibinfo {author} {\bibfnamefont {A.}~\bibnamefont
  {Abanov}}\ and\ \bibinfo {author} {\bibfnamefont {A.~V.}\ \bibnamefont
  {Chubukov}},\ }\href {https://doi.org/10.1103/PhysRevLett.84.5608} {\bibfield
   {journal} {\bibinfo  {journal} {Phys. Rev. Lett.}\ }\textbf {\bibinfo
  {volume} {84}},\ \bibinfo {pages} {5608} (\bibinfo {year}
  {2000})}\BibitemShut {NoStop}%
\bibitem [{\citenamefont {Abanov}\ \emph {et~al.}(2003)\citenamefont {Abanov},
  \citenamefont {Chubukov},\ and\ \citenamefont {Schmalian}}]{abanov2003}%
  \BibitemOpen
  \bibfield  {author} {\bibinfo {author} {\bibfnamefont {A.}~\bibnamefont
  {Abanov}}, \bibinfo {author} {\bibfnamefont {A.~V.}\ \bibnamefont
  {Chubukov}},\ and\ \bibinfo {author} {\bibfnamefont {J.}~\bibnamefont
  {Schmalian}},\ }\href {https://doi.org/10.1080/0001873021000057123}
  {\bibfield  {journal} {\bibinfo  {journal} {Advances in Physics}\ }\textbf
  {\bibinfo {volume} {52}},\ \bibinfo {pages} {119} (\bibinfo {year}
  {2003})}\BibitemShut {NoStop}%
\bibitem [{\citenamefont {Gelfand}\ \emph {et~al.}(1990)\citenamefont
  {Gelfand}, \citenamefont {Singh},\ and\ \citenamefont {Huse}}]{gelfand1990}%
  \BibitemOpen
  \bibfield  {author} {\bibinfo {author} {\bibfnamefont {M.~P.}\ \bibnamefont
  {Gelfand}}, \bibinfo {author} {\bibfnamefont {R.~R.}\ \bibnamefont {Singh}},\
  and\ \bibinfo {author} {\bibfnamefont {D.~A.}\ \bibnamefont {Huse}},\ }\href
  {https://doi.org/10.1007/BF01334744} {\bibfield  {journal} {\bibinfo
  {journal} {Journal of statistical Physics}\ }\textbf {\bibinfo {volume}
  {59}},\ \bibinfo {pages} {1093} (\bibinfo {year} {1990})}\BibitemShut
  {NoStop}%
\bibitem [{\citenamefont {Kubo}(1962)}]{kubo1962}%
  \BibitemOpen
  \bibfield  {author} {\bibinfo {author} {\bibfnamefont {R.}~\bibnamefont
  {Kubo}},\ }\href {https://doi.org/10.1143/JPSJ.17.1100} {\bibfield  {journal}
  {\bibinfo  {journal} {Journal of the Physical Society of Japan}\ }\textbf
  {\bibinfo {volume} {17}},\ \bibinfo {pages} {1100} (\bibinfo {year}
  {1962})}\BibitemShut {NoStop}%
\bibitem [{\citenamefont {Weihong}\ \emph {et~al.}(1994)\citenamefont
  {Weihong}, \citenamefont {Oitmaa},\ and\ \citenamefont
  {Hamer}}]{weihong1994}%
  \BibitemOpen
  \bibfield  {author} {\bibinfo {author} {\bibfnamefont {Z.}~\bibnamefont
  {Weihong}}, \bibinfo {author} {\bibfnamefont {J.}~\bibnamefont {Oitmaa}},\
  and\ \bibinfo {author} {\bibfnamefont {C.}~\bibnamefont {Hamer}},\ }\href
  {https://doi.org/10.1088/0305-4470/27/16/010} {\bibfield  {journal} {\bibinfo
   {journal} {Journal of Physics A: Mathematical and General}\ }\textbf
  {\bibinfo {volume} {27}},\ \bibinfo {pages} {5425} (\bibinfo {year}
  {1994})}\BibitemShut {NoStop}%
\bibitem [{\citenamefont {Nagai}\ \emph {et~al.}(1987)\citenamefont {Nagai},
  \citenamefont {Yamada}, \citenamefont {Nishino},\ and\ \citenamefont
  {Miyatake}}]{nagai1987}%
  \BibitemOpen
  \bibfield  {author} {\bibinfo {author} {\bibfnamefont {O.}~\bibnamefont
  {Nagai}}, \bibinfo {author} {\bibfnamefont {Y.}~\bibnamefont {Yamada}},
  \bibinfo {author} {\bibfnamefont {K.}~\bibnamefont {Nishino}},\ and\ \bibinfo
  {author} {\bibfnamefont {Y.}~\bibnamefont {Miyatake}},\ }\href
  {https://doi.org/10.1103/PhysRevB.35.3425} {\bibfield  {journal} {\bibinfo
  {journal} {Phys. Rev. B}\ }\textbf {\bibinfo {volume} {35}},\ \bibinfo
  {pages} {3425} (\bibinfo {year} {1987})}\BibitemShut {NoStop}%
\bibitem [{\citenamefont {Savit}(1980)}]{savit1980}%
  \BibitemOpen
  \bibfield  {author} {\bibinfo {author} {\bibfnamefont {R.}~\bibnamefont
  {Savit}},\ }\href {https://doi.org/10.1103/RevModPhys.52.453} {\bibfield
  {journal} {\bibinfo  {journal} {Rev. Mod. Phys.}\ }\textbf {\bibinfo {volume}
  {52}},\ \bibinfo {pages} {453} (\bibinfo {year} {1980})}\BibitemShut
  {NoStop}%
\bibitem [{\citenamefont {Gregor}\ \emph {et~al.}(2011)\citenamefont {Gregor},
  \citenamefont {Huse}, \citenamefont {Moessner},\ and\ \citenamefont
  {Sondhi}}]{gregor2011}%
  \BibitemOpen
  \bibfield  {author} {\bibinfo {author} {\bibfnamefont {K.}~\bibnamefont
  {Gregor}}, \bibinfo {author} {\bibfnamefont {D.~A.}\ \bibnamefont {Huse}},
  \bibinfo {author} {\bibfnamefont {R.}~\bibnamefont {Moessner}},\ and\
  \bibinfo {author} {\bibfnamefont {S.~L.}\ \bibnamefont {Sondhi}},\ }\href
  {https://doi.org/10.1088/1367-2630/13/2/025009} {\bibfield  {journal}
  {\bibinfo  {journal} {New Journal of Physics}\ }\textbf {\bibinfo {volume}
  {13}},\ \bibinfo {pages} {025009} (\bibinfo {year} {2011})}\BibitemShut
  {NoStop}%
\bibitem [{\citenamefont {Balian}\ \emph
  {et~al.}(1975{\natexlab{a}})\citenamefont {Balian}, \citenamefont {Drouffe},\
  and\ \citenamefont {Itzykson}}]{balian1975gauge2}%
  \BibitemOpen
  \bibfield  {author} {\bibinfo {author} {\bibfnamefont {R.}~\bibnamefont
  {Balian}}, \bibinfo {author} {\bibfnamefont {J.~M.}\ \bibnamefont
  {Drouffe}},\ and\ \bibinfo {author} {\bibfnamefont {C.}~\bibnamefont
  {Itzykson}},\ }\href {https://doi.org/10.1103/PhysRevD.11.2098} {\bibfield
  {journal} {\bibinfo  {journal} {Phys. Rev. D}\ }\textbf {\bibinfo {volume}
  {11}},\ \bibinfo {pages} {2098} (\bibinfo {year}
  {1975}{\natexlab{a}})}\BibitemShut {NoStop}%
\bibitem [{\citenamefont {Balian}\ \emph
  {et~al.}(1975{\natexlab{b}})\citenamefont {Balian}, \citenamefont {Drouffe},\
  and\ \citenamefont {Itzykson}}]{balian1975gauge3}%
  \BibitemOpen
  \bibfield  {author} {\bibinfo {author} {\bibfnamefont {R.}~\bibnamefont
  {Balian}}, \bibinfo {author} {\bibfnamefont {J.~M.}\ \bibnamefont
  {Drouffe}},\ and\ \bibinfo {author} {\bibfnamefont {C.}~\bibnamefont
  {Itzykson}},\ }\href {https://doi.org/10.1103/PhysRevD.11.2104} {\bibfield
  {journal} {\bibinfo  {journal} {Phys. Rev. D}\ }\textbf {\bibinfo {volume}
  {11}},\ \bibinfo {pages} {2104} (\bibinfo {year}
  {1975}{\natexlab{b}})}\BibitemShut {NoStop}%
\bibitem [{\citenamefont {Fradkin}\ and\ \citenamefont
  {Shenker}(1979)}]{fradkin1979}%
  \BibitemOpen
  \bibfield  {author} {\bibinfo {author} {\bibfnamefont {E.}~\bibnamefont
  {Fradkin}}\ and\ \bibinfo {author} {\bibfnamefont {S.~H.}\ \bibnamefont
  {Shenker}},\ }\href {https://doi.org/10.1103/PhysRevD.19.3682} {\bibfield
  {journal} {\bibinfo  {journal} {Phys. Rev. D}\ }\textbf {\bibinfo {volume}
  {19}},\ \bibinfo {pages} {3682} (\bibinfo {year} {1979})}\BibitemShut
  {NoStop}%
\bibitem [{\citenamefont {Tupitsyn}\ \emph {et~al.}(2010)\citenamefont
  {Tupitsyn}, \citenamefont {Kitaev}, \citenamefont {Prokof'ev},\ and\
  \citenamefont {Stamp}}]{tupitsyn2010}%
  \BibitemOpen
  \bibfield  {author} {\bibinfo {author} {\bibfnamefont {I.~S.}\ \bibnamefont
  {Tupitsyn}}, \bibinfo {author} {\bibfnamefont {A.}~\bibnamefont {Kitaev}},
  \bibinfo {author} {\bibfnamefont {N.~V.}\ \bibnamefont {Prokof'ev}},\ and\
  \bibinfo {author} {\bibfnamefont {P.~C.~E.}\ \bibnamefont {Stamp}},\ }\href
  {https://doi.org/10.1103/PhysRevB.82.085114} {\bibfield  {journal} {\bibinfo
  {journal} {Phys. Rev. B}\ }\textbf {\bibinfo {volume} {82}},\ \bibinfo
  {pages} {085114} (\bibinfo {year} {2010})}\BibitemShut {NoStop}%
\bibitem [{\citenamefont {Prosko}\ \emph {et~al.}(2017)\citenamefont {Prosko},
  \citenamefont {Lee},\ and\ \citenamefont {Maciejko}}]{prosko2017}%
  \BibitemOpen
  \bibfield  {author} {\bibinfo {author} {\bibfnamefont {C.}~\bibnamefont
  {Prosko}}, \bibinfo {author} {\bibfnamefont {S.-P.}\ \bibnamefont {Lee}},\
  and\ \bibinfo {author} {\bibfnamefont {J.}~\bibnamefont {Maciejko}},\ }\href
  {https://doi.org/10.1103/PhysRevB.96.205104} {\bibfield  {journal} {\bibinfo
  {journal} {Phys. Rev. B}\ }\textbf {\bibinfo {volume} {96}},\ \bibinfo
  {pages} {205104} (\bibinfo {year} {2017})}\BibitemShut {NoStop}%
\bibitem [{\citenamefont {Wegner}(1971)}]{wegner1971}%
  \BibitemOpen
  \bibfield  {author} {\bibinfo {author} {\bibfnamefont {F.~J.}\ \bibnamefont
  {Wegner}},\ }\href {https://doi.org/10.1063/1.1665530} {\bibfield  {journal}
  {\bibinfo  {journal} {Journal of Mathematical Physics}\ }\textbf {\bibinfo
  {volume} {12}},\ \bibinfo {pages} {2259} (\bibinfo {year}
  {1971})}\BibitemShut {NoStop}%
\bibitem [{\citenamefont {Qin}\ \emph {et~al.}(2017)\citenamefont {Qin},
  \citenamefont {He}, \citenamefont {You}, \citenamefont {Lu}, \citenamefont
  {Sen}, \citenamefont {Sandvik}, \citenamefont {Xu},\ and\ \citenamefont
  {Meng}}]{qin2017}%
  \BibitemOpen
  \bibfield  {author} {\bibinfo {author} {\bibfnamefont {Y.~Q.}\ \bibnamefont
  {Qin}}, \bibinfo {author} {\bibfnamefont {Y.-Y.}\ \bibnamefont {He}},
  \bibinfo {author} {\bibfnamefont {Y.-Z.}\ \bibnamefont {You}}, \bibinfo
  {author} {\bibfnamefont {Z.-Y.}\ \bibnamefont {Lu}}, \bibinfo {author}
  {\bibfnamefont {A.}~\bibnamefont {Sen}}, \bibinfo {author} {\bibfnamefont
  {A.~W.}\ \bibnamefont {Sandvik}}, \bibinfo {author} {\bibfnamefont
  {C.}~\bibnamefont {Xu}},\ and\ \bibinfo {author} {\bibfnamefont {Z.~Y.}\
  \bibnamefont {Meng}},\ }\href {https://doi.org/10.1103/PhysRevX.7.031052}
  {\bibfield  {journal} {\bibinfo  {journal} {Phys. Rev. X}\ }\textbf {\bibinfo
  {volume} {7}},\ \bibinfo {pages} {031052} (\bibinfo {year}
  {2017})}\BibitemShut {NoStop}%
\bibitem [{\citenamefont {Senthil}\ \emph {et~al.}(2019)\citenamefont
  {Senthil}, \citenamefont {Son}, \citenamefont {Wang},\ and\ \citenamefont
  {Xu}}]{senthil2019}%
  \BibitemOpen
  \bibfield  {author} {\bibinfo {author} {\bibfnamefont {T.}~\bibnamefont
  {Senthil}}, \bibinfo {author} {\bibfnamefont {D.~T.}\ \bibnamefont {Son}},
  \bibinfo {author} {\bibfnamefont {C.}~\bibnamefont {Wang}},\ and\ \bibinfo
  {author} {\bibfnamefont {C.}~\bibnamefont {Xu}},\ }\href
  {https://doi.org/10.1016/j.physrep.2019.09.001} {\bibfield  {journal}
  {\bibinfo  {journal} {Physics Reports}\ }\textbf {\bibinfo {volume} {827}},\
  \bibinfo {pages} {1} (\bibinfo {year} {2019})}\BibitemShut {NoStop}%
\bibitem [{\citenamefont {Jian}\ \emph {et~al.}(2019)\citenamefont {Jian},
  \citenamefont {Bi},\ and\ \citenamefont {You}}]{jian2019}%
  \BibitemOpen
  \bibfield  {author} {\bibinfo {author} {\bibfnamefont {C.-M.}\ \bibnamefont
  {Jian}}, \bibinfo {author} {\bibfnamefont {Z.}~\bibnamefont {Bi}},\ and\
  \bibinfo {author} {\bibfnamefont {Y.-Z.}\ \bibnamefont {You}},\ }\href
  {https://doi.org/10.1103/PhysRevB.100.075109} {\bibfield  {journal} {\bibinfo
   {journal} {Phys. Rev. B}\ }\textbf {\bibinfo {volume} {100}},\ \bibinfo
  {pages} {075109} (\bibinfo {year} {2019})}\BibitemShut {NoStop}%
\bibitem [{\citenamefont {Chubukov}\ and\ \citenamefont
  {Maslov}(2003)}]{chubukov2003}%
  \BibitemOpen
  \bibfield  {author} {\bibinfo {author} {\bibfnamefont {A.~V.}\ \bibnamefont
  {Chubukov}}\ and\ \bibinfo {author} {\bibfnamefont {D.~L.}\ \bibnamefont
  {Maslov}},\ }\href {https://doi.org/10.1103/PhysRevB.68.155113} {\bibfield
  {journal} {\bibinfo  {journal} {Phys. Rev. B}\ }\textbf {\bibinfo {volume}
  {68}},\ \bibinfo {pages} {155113} (\bibinfo {year} {2003})}\BibitemShut
  {NoStop}%
\bibitem [{\citenamefont {Belitz}\ \emph {et~al.}(2005)\citenamefont {Belitz},
  \citenamefont {Kirkpatrick},\ and\ \citenamefont {Vojta}}]{belitz2005}%
  \BibitemOpen
  \bibfield  {author} {\bibinfo {author} {\bibfnamefont {D.}~\bibnamefont
  {Belitz}}, \bibinfo {author} {\bibfnamefont {T.~R.}\ \bibnamefont
  {Kirkpatrick}},\ and\ \bibinfo {author} {\bibfnamefont {T.}~\bibnamefont
  {Vojta}},\ }\href {https://doi.org/10.1103/RevModPhys.77.579} {\bibfield
  {journal} {\bibinfo  {journal} {Rev. Mod. Phys.}\ }\textbf {\bibinfo {volume}
  {77}},\ \bibinfo {pages} {579} (\bibinfo {year} {2005})}\BibitemShut
  {NoStop}%
\bibitem [{\citenamefont {Ferrenberg}\ \emph {et~al.}(2018)\citenamefont
  {Ferrenberg}, \citenamefont {Xu},\ and\ \citenamefont
  {Landau}}]{ferrenberg2018}%
  \BibitemOpen
  \bibfield  {author} {\bibinfo {author} {\bibfnamefont {A.~M.}\ \bibnamefont
  {Ferrenberg}}, \bibinfo {author} {\bibfnamefont {J.}~\bibnamefont {Xu}},\
  and\ \bibinfo {author} {\bibfnamefont {D.~P.}\ \bibnamefont {Landau}},\
  }\href {https://doi.org/10.1103/PhysRevE.97.043301} {\bibfield  {journal}
  {\bibinfo  {journal} {Phys. Rev. E}\ }\textbf {\bibinfo {volume} {97}},\
  \bibinfo {pages} {043301} (\bibinfo {year} {2018})}\BibitemShut {NoStop}%
\bibitem [{\citenamefont {de'Medici}\ \emph {et~al.}(2005)\citenamefont
  {de'Medici}, \citenamefont {Georges},\ and\ \citenamefont
  {Biermann}}]{demedici2005}%
  \BibitemOpen
  \bibfield  {author} {\bibinfo {author} {\bibfnamefont {L.}~\bibnamefont
  {de'Medici}}, \bibinfo {author} {\bibfnamefont {A.}~\bibnamefont {Georges}},\
  and\ \bibinfo {author} {\bibfnamefont {S.}~\bibnamefont {Biermann}},\ }\href
  {https://doi.org/10.1103/PhysRevB.72.205124} {\bibfield  {journal} {\bibinfo
  {journal} {Phys. Rev. B}\ }\textbf {\bibinfo {volume} {72}},\ \bibinfo
  {pages} {205124} (\bibinfo {year} {2005})}\BibitemShut {NoStop}%
\bibitem [{\citenamefont {R\"uegg}\ \emph {et~al.}(2010)\citenamefont
  {R\"uegg}, \citenamefont {Huber},\ and\ \citenamefont {Sigrist}}]{ruegg2010}%
  \BibitemOpen
  \bibfield  {author} {\bibinfo {author} {\bibfnamefont {A.}~\bibnamefont
  {R\"uegg}}, \bibinfo {author} {\bibfnamefont {S.~D.}\ \bibnamefont {Huber}},\
  and\ \bibinfo {author} {\bibfnamefont {M.}~\bibnamefont {Sigrist}},\ }\href
  {https://doi.org/10.1103/PhysRevB.81.155118} {\bibfield  {journal} {\bibinfo
  {journal} {Phys. Rev. B}\ }\textbf {\bibinfo {volume} {81}},\ \bibinfo
  {pages} {155118} (\bibinfo {year} {2010})}\BibitemShut {NoStop}%
\bibitem [{\citenamefont {Hassan}\ and\ \citenamefont {de'
  Medici}(2010)}]{hassan2010}%
  \BibitemOpen
  \bibfield  {author} {\bibinfo {author} {\bibfnamefont {S.~R.}\ \bibnamefont
  {Hassan}}\ and\ \bibinfo {author} {\bibfnamefont {L.}~\bibnamefont {de'
  Medici}},\ }\href {https://doi.org/10.1103/PhysRevB.81.035106} {\bibfield
  {journal} {\bibinfo  {journal} {Phys. Rev. B}\ }\textbf {\bibinfo {volume}
  {81}},\ \bibinfo {pages} {035106} (\bibinfo {year} {2010})}\BibitemShut
  {NoStop}%
\bibitem [{\citenamefont {Yu}\ and\ \citenamefont {Si}(2011)}]{yu2011}%
  \BibitemOpen
  \bibfield  {author} {\bibinfo {author} {\bibfnamefont {R.}~\bibnamefont
  {Yu}}\ and\ \bibinfo {author} {\bibfnamefont {Q.}~\bibnamefont {Si}},\ }\href
  {https://doi.org/10.1103/PhysRevB.84.235115} {\bibfield  {journal} {\bibinfo
  {journal} {Phys. Rev. B}\ }\textbf {\bibinfo {volume} {84}},\ \bibinfo
  {pages} {235115} (\bibinfo {year} {2011})}\BibitemShut {NoStop}%
\bibitem [{\citenamefont {Zhong}\ \emph {et~al.}(2012)\citenamefont {Zhong},
  \citenamefont {Liu}, \citenamefont {Wang},\ and\ \citenamefont
  {Luo}}]{zhong2012}%
  \BibitemOpen
  \bibfield  {author} {\bibinfo {author} {\bibfnamefont {Y.}~\bibnamefont
  {Zhong}}, \bibinfo {author} {\bibfnamefont {K.}~\bibnamefont {Liu}}, \bibinfo
  {author} {\bibfnamefont {Y.-Q.}\ \bibnamefont {Wang}},\ and\ \bibinfo
  {author} {\bibfnamefont {H.-G.}\ \bibnamefont {Luo}},\ }\href
  {https://doi.org/10.1103/PhysRevB.86.165134} {\bibfield  {journal} {\bibinfo
  {journal} {Phys. Rev. B}\ }\textbf {\bibinfo {volume} {86}},\ \bibinfo
  {pages} {165134} (\bibinfo {year} {2012})}\BibitemShut {NoStop}%
\bibitem [{\citenamefont {Zhong}\ and\ \citenamefont {Luo}(2013)}]{zhong2013}%
  \BibitemOpen
  \bibfield  {author} {\bibinfo {author} {\bibfnamefont {Y.}~\bibnamefont
  {Zhong}}\ and\ \bibinfo {author} {\bibfnamefont {H.-G.}\ \bibnamefont
  {Luo}},\ }\href {https://doi.org/10.1142/S021797921361002X} {\bibfield
  {journal} {\bibinfo  {journal} {International Journal of Modern Physics B}\
  }\textbf {\bibinfo {volume} {27}},\ \bibinfo {pages} {1361002} (\bibinfo
  {year} {2013})}\BibitemShut {NoStop}%
\bibitem [{\citenamefont {Gazit}\ \emph {et~al.}(2020)\citenamefont {Gazit},
  \citenamefont {Assaad},\ and\ \citenamefont {Sachdev}}]{gazit2020}%
  \BibitemOpen
  \bibfield  {author} {\bibinfo {author} {\bibfnamefont {S.}~\bibnamefont
  {Gazit}}, \bibinfo {author} {\bibfnamefont {F.~F.}\ \bibnamefont {Assaad}},\
  and\ \bibinfo {author} {\bibfnamefont {S.}~\bibnamefont {Sachdev}},\ }\href
  {https://doi.org/10.1103/PhysRevX.10.041057} {\bibfield  {journal} {\bibinfo
  {journal} {Phys. Rev. X}\ }\textbf {\bibinfo {volume} {10}},\ \bibinfo
  {pages} {041057} (\bibinfo {year} {2020})}\BibitemShut {NoStop}%
\bibitem [{\citenamefont {Damascelli}\ \emph {et~al.}(2003)\citenamefont
  {Damascelli}, \citenamefont {Hussain},\ and\ \citenamefont
  {Shen}}]{damascelli2003}%
  \BibitemOpen
  \bibfield  {author} {\bibinfo {author} {\bibfnamefont {A.}~\bibnamefont
  {Damascelli}}, \bibinfo {author} {\bibfnamefont {Z.}~\bibnamefont
  {Hussain}},\ and\ \bibinfo {author} {\bibfnamefont {Z.-X.}\ \bibnamefont
  {Shen}},\ }\href {https://doi.org/10.1103/RevModPhys.75.473} {\bibfield
  {journal} {\bibinfo  {journal} {Rev. Mod. Phys.}\ }\textbf {\bibinfo {volume}
  {75}},\ \bibinfo {pages} {473} (\bibinfo {year} {2003})}\BibitemShut
  {NoStop}%
\bibitem [{\citenamefont {Sobota}\ \emph {et~al.}(2021)\citenamefont {Sobota},
  \citenamefont {He},\ and\ \citenamefont {Shen}}]{sobota2021}%
  \BibitemOpen
  \bibfield  {author} {\bibinfo {author} {\bibfnamefont {J.~A.}\ \bibnamefont
  {Sobota}}, \bibinfo {author} {\bibfnamefont {Y.}~\bibnamefont {He}},\ and\
  \bibinfo {author} {\bibfnamefont {Z.-X.}\ \bibnamefont {Shen}},\ }\href
  {https://doi.org/10.1103/RevModPhys.93.025006} {\bibfield  {journal}
  {\bibinfo  {journal} {Rev. Mod. Phys.}\ }\textbf {\bibinfo {volume} {93}},\
  \bibinfo {pages} {025006} (\bibinfo {year} {2021})}\BibitemShut {NoStop}%
\bibitem [{\citenamefont {Ok}\ \emph {et~al.}(2017)\citenamefont {Ok},
  \citenamefont {Baek}, \citenamefont {Hoch}, \citenamefont {Kremer},
  \citenamefont {Park}, \citenamefont {Ji}, \citenamefont {B{\"u}chner},
  \citenamefont {Park}, \citenamefont {Hyun}, \citenamefont {Shim} \emph
  {et~al.}}]{ok2017}%
  \BibitemOpen
  \bibfield  {author} {\bibinfo {author} {\bibfnamefont {J.~M.}\ \bibnamefont
  {Ok}}, \bibinfo {author} {\bibfnamefont {S.-H.}\ \bibnamefont {Baek}},
  \bibinfo {author} {\bibfnamefont {C.}~\bibnamefont {Hoch}}, \bibinfo {author}
  {\bibfnamefont {R.}~\bibnamefont {Kremer}}, \bibinfo {author} {\bibfnamefont
  {S.}~\bibnamefont {Park}}, \bibinfo {author} {\bibfnamefont {S.}~\bibnamefont
  {Ji}}, \bibinfo {author} {\bibfnamefont {B.}~\bibnamefont {B{\"u}chner}},
  \bibinfo {author} {\bibfnamefont {J.-H.}\ \bibnamefont {Park}}, \bibinfo
  {author} {\bibfnamefont {S.}~\bibnamefont {Hyun}}, \bibinfo {author}
  {\bibfnamefont {J.}~\bibnamefont {Shim}}, \emph {et~al.},\ }\href
  {https://doi.org/10.1038/s41467-017-02327-0} {\bibfield  {journal} {\bibinfo
  {journal} {Nature communications}\ }\textbf {\bibinfo {volume} {8}},\
  \bibinfo {pages} {2167} (\bibinfo {year} {2017})}\BibitemShut {NoStop}%
\bibitem [{\citenamefont {Lee}\ \emph {et~al.}(2019)\citenamefont {Lee},
  \citenamefont {Roh}, \citenamefont {Seo}, \citenamefont {Lee}, \citenamefont
  {Jung}, \citenamefont {Mok~Ok}, \citenamefont {Jung}, \citenamefont {Kang},
  \citenamefont {Lee}, \citenamefont {Sung~Kim} \emph {et~al.}}]{lee2019}%
  \BibitemOpen
  \bibfield  {author} {\bibinfo {author} {\bibfnamefont {S.}~\bibnamefont
  {Lee}}, \bibinfo {author} {\bibfnamefont {S.}~\bibnamefont {Roh}}, \bibinfo
  {author} {\bibfnamefont {Y.-S.}\ \bibnamefont {Seo}}, \bibinfo {author}
  {\bibfnamefont {M.}~\bibnamefont {Lee}}, \bibinfo {author} {\bibfnamefont
  {E.}~\bibnamefont {Jung}}, \bibinfo {author} {\bibfnamefont {J.}~\bibnamefont
  {Mok~Ok}}, \bibinfo {author} {\bibfnamefont {M.-C.}\ \bibnamefont {Jung}},
  \bibinfo {author} {\bibfnamefont {B.}~\bibnamefont {Kang}}, \bibinfo {author}
  {\bibfnamefont {K.-W.}\ \bibnamefont {Lee}}, \bibinfo {author} {\bibfnamefont
  {J.}~\bibnamefont {Sung~Kim}}, \emph {et~al.},\ }\href
  {https://doi.org/10.1088/1361-648X/ab31cd} {\bibfield  {journal} {\bibinfo
  {journal} {Journal of Physics: Condensed Matter}\ }\textbf {\bibinfo {volume}
  {31}},\ \bibinfo {pages} {445602} (\bibinfo {year} {2019})}\BibitemShut
  {NoStop}%
\bibitem [{\citenamefont {Choi}\ \emph {et~al.}(2017)\citenamefont {Choi},
  \citenamefont {Choi}, \citenamefont {Ok}, \citenamefont {Lee}, \citenamefont
  {Jang}, \citenamefont {Lee}, \citenamefont {Kuk}, \citenamefont {Lee},
  \citenamefont {Heinrich}, \citenamefont {Cheong}, \citenamefont {Bang},
  \citenamefont {Johnston}, \citenamefont {Kim},\ and\ \citenamefont
  {Lee}}]{choi2017}%
  \BibitemOpen
  \bibfield  {author} {\bibinfo {author} {\bibfnamefont {S.}~\bibnamefont
  {Choi}}, \bibinfo {author} {\bibfnamefont {H.~J.}\ \bibnamefont {Choi}},
  \bibinfo {author} {\bibfnamefont {J.~M.}\ \bibnamefont {Ok}}, \bibinfo
  {author} {\bibfnamefont {Y.}~\bibnamefont {Lee}}, \bibinfo {author}
  {\bibfnamefont {W.-J.}\ \bibnamefont {Jang}}, \bibinfo {author}
  {\bibfnamefont {A.~T.}\ \bibnamefont {Lee}}, \bibinfo {author} {\bibfnamefont
  {Y.}~\bibnamefont {Kuk}}, \bibinfo {author} {\bibfnamefont {S.}~\bibnamefont
  {Lee}}, \bibinfo {author} {\bibfnamefont {A.~J.}\ \bibnamefont {Heinrich}},
  \bibinfo {author} {\bibfnamefont {S.-W.}\ \bibnamefont {Cheong}}, \bibinfo
  {author} {\bibfnamefont {Y.}~\bibnamefont {Bang}}, \bibinfo {author}
  {\bibfnamefont {S.}~\bibnamefont {Johnston}}, \bibinfo {author}
  {\bibfnamefont {J.~S.}\ \bibnamefont {Kim}},\ and\ \bibinfo {author}
  {\bibfnamefont {J.}~\bibnamefont {Lee}},\ }\href
  {https://doi.org/10.1103/PhysRevLett.119.227001} {\bibfield  {journal}
  {\bibinfo  {journal} {Phys. Rev. Lett.}\ }\textbf {\bibinfo {volume} {119}},\
  \bibinfo {pages} {227001} (\bibinfo {year} {2017})}\BibitemShut {NoStop}%
\bibitem [{\citenamefont {Kim}\ \emph {et~al.}(2021)\citenamefont {Kim},
  \citenamefont {Ok}, \citenamefont {Oh}, \citenamefont {Kwon}, \citenamefont
  {Zhang}, \citenamefont {Denlinger}, \citenamefont {Mo}, \citenamefont
  {Wolff-Fabris}, \citenamefont {Kampert}, \citenamefont {Moon} \emph
  {et~al.}}]{kim2021}%
  \BibitemOpen
  \bibfield  {author} {\bibinfo {author} {\bibfnamefont {S.}~\bibnamefont
  {Kim}}, \bibinfo {author} {\bibfnamefont {J.~M.}\ \bibnamefont {Ok}},
  \bibinfo {author} {\bibfnamefont {H.}~\bibnamefont {Oh}}, \bibinfo {author}
  {\bibfnamefont {C.~I.}\ \bibnamefont {Kwon}}, \bibinfo {author}
  {\bibfnamefont {Y.}~\bibnamefont {Zhang}}, \bibinfo {author} {\bibfnamefont
  {J.~D.}\ \bibnamefont {Denlinger}}, \bibinfo {author} {\bibfnamefont {S.-K.}\
  \bibnamefont {Mo}}, \bibinfo {author} {\bibfnamefont {F.}~\bibnamefont
  {Wolff-Fabris}}, \bibinfo {author} {\bibfnamefont {E.}~\bibnamefont
  {Kampert}}, \bibinfo {author} {\bibfnamefont {E.-G.}\ \bibnamefont {Moon}},
  \emph {et~al.},\ }\href {https://doi.org/10.1073/pnas.2105190118} {\bibfield
  {journal} {\bibinfo  {journal} {Proceedings of the National Academy of
  Sciences}\ }\textbf {\bibinfo {volume} {118}},\ \bibinfo {pages}
  {e2105190118} (\bibinfo {year} {2021})}\BibitemShut {NoStop}%
\bibitem [{\citenamefont {Qian}\ \emph {et~al.}(2011)\citenamefont {Qian},
  \citenamefont {Xu}, \citenamefont {Shi}, \citenamefont {Nakayama},
  \citenamefont {Richard}, \citenamefont {Kawahara}, \citenamefont {Sato},
  \citenamefont {Takahashi}, \citenamefont {Neupane}, \citenamefont {Xu},
  \citenamefont {Wang}, \citenamefont {Xu}, \citenamefont {Dai}, \citenamefont
  {Fang}, \citenamefont {Cheng}, \citenamefont {Wen},\ and\ \citenamefont
  {Ding}}]{qian2011}%
  \BibitemOpen
  \bibfield  {author} {\bibinfo {author} {\bibfnamefont {T.}~\bibnamefont
  {Qian}}, \bibinfo {author} {\bibfnamefont {N.}~\bibnamefont {Xu}}, \bibinfo
  {author} {\bibfnamefont {Y.-B.}\ \bibnamefont {Shi}}, \bibinfo {author}
  {\bibfnamefont {K.}~\bibnamefont {Nakayama}}, \bibinfo {author}
  {\bibfnamefont {P.}~\bibnamefont {Richard}}, \bibinfo {author} {\bibfnamefont
  {T.}~\bibnamefont {Kawahara}}, \bibinfo {author} {\bibfnamefont
  {T.}~\bibnamefont {Sato}}, \bibinfo {author} {\bibfnamefont {T.}~\bibnamefont
  {Takahashi}}, \bibinfo {author} {\bibfnamefont {M.}~\bibnamefont {Neupane}},
  \bibinfo {author} {\bibfnamefont {Y.-M.}\ \bibnamefont {Xu}}, \bibinfo
  {author} {\bibfnamefont {X.-P.}\ \bibnamefont {Wang}}, \bibinfo {author}
  {\bibfnamefont {G.}~\bibnamefont {Xu}}, \bibinfo {author} {\bibfnamefont
  {X.}~\bibnamefont {Dai}}, \bibinfo {author} {\bibfnamefont {Z.}~\bibnamefont
  {Fang}}, \bibinfo {author} {\bibfnamefont {P.}~\bibnamefont {Cheng}},
  \bibinfo {author} {\bibfnamefont {H.-H.}\ \bibnamefont {Wen}},\ and\ \bibinfo
  {author} {\bibfnamefont {H.}~\bibnamefont {Ding}},\ }\href
  {https://doi.org/10.1103/PhysRevB.83.140513} {\bibfield  {journal} {\bibinfo
  {journal} {Phys. Rev. B}\ }\textbf {\bibinfo {volume} {83}},\ \bibinfo
  {pages} {140513(R)} (\bibinfo {year} {2011})}\BibitemShut {NoStop}%
\end{thebibliography}%

\onecolumngrid
\clearpage

\begin{center}
    {\large\bfseries Supplemental Material for}\\[4pt]
    {\large\bfseries
    \textquotedblleft Thermal Phase Transition of Three Dimensional Orthogonal Metals\textquotedblright}
\end{center}
\vspace{0.5cm}
\begin{center}
Inho Song,$^{1,\textcolor{red}{*}}$ Minsu Park,$^{1,\textcolor{red}{*}}$ Hanbit Oh,$^{2,\textcolor{red}{\dagger}}$ and Eun-Gook Moon$^{1,\textcolor{red}{\ddagger}}$
\vspace{0.2cm}

{\small
$^{1}$Department of Physics, Korea Advanced Institute of Science and
Technology (KAIST), Daejeon 34141, Korea\\
$^{2}$Department of Physics, University of Wisconsin-Milwaukee, Milwaukee, Wisconsin 53201, USA\\
}
\end{center}
\vspace{0.5cm}
\setcounter{page}{1}

\setcounter{section}{0}
\setcounter{subsection}{0}
\setcounter{equation}{0}
\setcounter{figure}{0}
\setcounter{table}{0}
\renewcommand{\thesection}{S\arabic{section}}
\renewcommand{\thesubsection}{S\arabic{section}.\arabic{subsection}}
\renewcommand{\theequation}{S\arabic{equation}}
\renewcommand{\thefigure}{S\arabic{figure}}
\renewcommand{\thetable}{S\arabic{table}}
\renewcommand{\theHsection}{supp.\arabic{section}}
\renewcommand{\theHsubsection}{supp.\arabic{section}.\arabic{subsection}}
\renewcommand{\theHequation}{S\arabic{equation}}
\renewcommand{\theHfigure}{S\arabic{figure}}
\renewcommand{\theHtable}{S\arabic{table}}

\let\addtocontents\SMsavedaddtocontents
\makeatletter
\def\tocmax@pagenum{2em}
\makeatother
\tableofcontents

\vspace{0.3cm}

\section{Shift of confinement temperature}
\label{sec:s1}

Consider the commuting star and plaquette Hamiltonian perturbed by a longitudinal field,
\begin{equation}
\label{eq:gauge_model}
H_0=-J_B\sum_p B_p-J_A\sum_r A_r,\quad
V=-h_z\sum_\ell\sigma_\ell^z.
\end{equation}
Here $B_p=\prod_{\ell \in\partial p}
\sigma_{\ell}^z$ and $A_r = \prod_{\ell \ni r}\sigma^x_{\ell}$. A single $\sigma_\ell^z$ anticommutes with the stars at the two endpoints of $\ell$, but commutes with every plaquette operator. An insertion therefore changes the star configuration during an imaginary-time interval without changing the conserved flux labels. A nonzero thermal trace requires the star history to close. At second order, inserting the same link twice restores the stars and produces a flux-independent contribution. Four distinct links around an elementary square instead restore every star while leaving the operator product $B_p$.

With $K_A=\beta J_A$, $K_B=\beta J_B$, $K_z=\beta h_z$, and $O=\sum_\ell\sigma_\ell^z$, the connected expansion is \citeS{S_kubo1962,S_gelfand1990}
\begin{equation}
\label{eq:cumulant}
\ln\frac{Z}{Z_0}
=\sum_{n\geq1}\frac{K_z^n}{n!}
\int_0^1dv_1\cdots\int_0^1dv_n\,
\langle T_v O(v_1)\cdots O(v_n)\rangle_{0,c}.
\end{equation}

The contribution involving the four distinct boundary links of one plaquette has no lower-order disconnected subcluster with the same flux dependence. Its star trace can therefore be written as
\begin{equation}
\label{eq:star_kernel}
\left\langle T_v\prod_{\ell\in\partial p}\sigma_\ell^z(v_\ell)\right\rangle_A
=C_A(\{v_\ell\};K_A)B_p.
\end{equation}

The task is to evaluate $C_A$, including the different chronological assignments of links to insertion times. Lower-order analytic terms are retained in the regular free energy but do not provide the leading flux-sensitive shift discussed here.

Label the plaquette vertices cyclically by $v_1,v_2,v_3,v_4$, and its edges by $e_0,e_1,e_2,e_3$, with $\partial e_0=\{v_1,v_2\}$, $\partial e_1=\{v_2,v_3\}$, $\partial e_2=\{v_3,v_4\}$, and $\partial e_3=\{v_4,v_1\}$. These are fixed spatial labels. Let $0<t_1<t_2<t_3<t_4<1$ be the ordered dimensionless insertion times and $\pi=(\pi_1,\pi_2,\pi_3,\pi_4)$ the corresponding edge ordering. After the first $j$ insertions, the vertices whose stars have been flipped an odd number of times form the set
\begin{equation}
\label{eq:flipped_set}
F_j(\pi)=\partial e_{\pi_1}\triangle\cdots\triangle\partial e_{\pi_j},
\qquad F_4(\pi)=\varnothing.
\end{equation}
The symmetric difference implements the fact that two flips cancel. For example, the ordering $(0,1,2,3)$ gives $F_1=\{v_1,v_2\}$, $F_2=\{v_1,v_3\}$, and $F_3=\{v_1,v_4\}$. Inserting opposite edges first, as in $(0,2,1,3)$, instead produces an intermediate state with all four stars flipped.

Let $A_v|a\rangle=a_v|a\rangle$, with $a_v=\pm1$, and define $d_j=t_{j+1}-t_j$ and $D=d_1+d_2+d_3$. The initial star energy is $E_0(a)=-J_A\sum_v a_v$. During interval $d_j$, its change is
\begin{equation}
\label{eq:history_energy}
E_j(a;\pi)-E_0(a)=2J_A\sum_{v\in F_j(\pi)}a_v.
\end{equation}

Inserting complete star eigenstates in the ordered trace gives
\begin{equation}
\label{eq:thermal_star_sum}
C_A^{(\pi)}
=\sum_{\{a_v\}}P_{K_A}(a)
\exp\!\left[-2K_A\sum_{j=1}^3d_j\sum_{v\in F_j(\pi)}a_v\right],
\quad
P_{K_A}(a)=\frac{\exp(K_A\sum_{v=1}^4a_v)}{(2\cosh K_A)^4}.
\end{equation}
The independent local star weights used here are the thermodynamic-limit weights; global star constraints on a finite periodic system do not change this finite local cluster in that limit. The factor before and after the four insertions is precisely the thermal probability $P_{K_A}$, while the exponential records the extra energy accumulated along the history.

For each vertex, introduce its total flipped duration,
\begin{equation}
\label{eq:duration}
\ell_v^{(\pi)}=\sum_{j:\,v\in F_j(\pi)}d_j.
\end{equation}
The exponent in Eq. \eqref{eq:thermal_star_sum} is then a sum of four separate vertex contributions. Summing each $a_v$ independently yields
\begin{equation}
\label{eq:factorized_kernel}
C_A^{(\pi)}(d_1,d_2,d_3;K_A)
=\prod_{v=1}^4g_{K_A}\!\left(\ell_v^{(\pi)}\right),
\qquad
g_K(\ell)=\frac{\cosh[K(1-2\ell)]}{\cosh K}.
\end{equation}
Thus the temporal kernel is the product of four thermal star propagators. This form includes both initially satisfied and initially excited stars and is valid at arbitrary $K_A$, rather than only in a zero-temperature star background.

Rotations and reflections of the square partition its $4!=24$ edge orderings into three classes, each containing eight members. Table \ref{tab:s1} lists their flipped durations. The first two classes both have two flipped stars at every intermediate step, but differ in which vertices remain flipped across successive intervals; this distinction matters at finite $K_A$.

\begin{table}[tb]
\centering
\small
\setlength{\tabcolsep}{5pt}
\renewcommand{\arraystretch}{1.25}
\begin{tabular}{@{}cccr@{}}
\hline
\hline
Ordering $\pi$ & $(n_1,n_2,n_3)$ & $(\ell_{v_1},\ell_{v_2},\ell_{v_3},\ell_{v_4})$ & Multiplicity \\
\hline
$(0,1,2,3)$ & $(2,2,2)$ & $(D,d_1,d_2,d_3)$ & 8 \\
$(0,1,3,2)$ & $(2,2,2)$ & $(d_1+d_2,d_1,d_2+d_3,d_3)$ & 8 \\
$(0,2,1,3)$ & $(2,4,2)$ & $(D,d_1+d_2,d_2,d_2+d_3)$ & 8 \\
\hline
\hline
\end{tabular}
\caption{Representative chronological edge orderings and vertex-resolved flipped durations. In every row the fourth insertion closes the history. The entries $n_j=|F_j|$ count intermediate flipped stars.}
\label{tab:s1}
\end{table}
The common starting time $t_1$ does not enter the kernel. At fixed $d_1,d_2,d_3$ its range is $0<t_1<1-D$, so integrating it out supplies the factor $1-D$. If $ \int_{\mathcal S}= \int_{\{d_j\geq0,\ D\leq1\}} \mathrm{d} d_1 \mathrm{d} d_2 \mathrm{d} d_3 $, the representative integrals are
\begin{align}
\label{eq:I0123}
I_{0123}(K)&=\int_{\mathcal S}  (1-D)
g_K(D)g_K(d_1)g_K(d_2)g_K(d_3),\\
\label{eq:I0132}
I_{0132}(K)&=\int_{\mathcal S}  (1-D)
g_K(d_1+d_2)g_K(d_1)g_K(d_2+d_3)g_K(d_3),\\
\label{eq:I0213}
I_{0213}(K)&=\int_{\mathcal S}  (1-D)
g_K(D)g_K(d_1+d_2)g_K(d_2)g_K(d_2+d_3),
\end{align}

Collecting the orderings gives the coefficient used in the main text,
\begin{equation}
\label{eq:c4}
c_4(K)=8\,[I_{0123}(K)+I_{0132}(K)+I_{0213}(K)].
\end{equation}
Equivalently, $c_4$ is the sum of $C_A^{(\pi)}$ integrated over the ordered four-time simplex. This ordered-domain convention already includes the chronological multiplicities; an additional factor $1/4!$ must not be applied to Eq. \eqref{eq:c4}. In the full-domain cumulant expansion, the same factor is accounted for by the assignments of the four distinct link operators.

At $J_A=0$, $g_0(\ell)=1$. Each integral is then the volume of the ordered four-time simplex, giving the exact normalization check
\begin{equation}
\label{eq:c4_check}
I_{0123}(0)=I_{0132}(0)=I_{0213}(0)=\frac1{24},
\qquad c_4(0)=1.
\end{equation}
For $K\geq0$ and $0\leq\ell\leq1$, $0<g_K(\ell)\leq1$, so the coefficient remains finite and positive. Its dependence on $J_A$ comes from the energy cost of the intermediate star histories.

The flux-sensitive part of the fourth-order free-energy density follows directly from $f=-T\ln Z/N$:
\begin{equation}
\label{eq:free_energy}
\delta f_{\mathrm{sing}}^{(4)}
=-T K_z^4c_4(K_A)
\left[\frac1N\sum_p\langle B_p\rangle_0\right]_{\mathrm{sing}}.
\end{equation}
Comparing with a change of the dimensionless plaquette coupling gives
\begin{equation}
\label{eq:Keff}
K_B^{\mathrm{eff}}=K_B+c_4(K_A)K_z^4+O(K_z^6).
\end{equation}

Let the confinement temperature at $h_z=0$ as $T_{\mathrm{conf}}=\kappa_{\mathrm{conf}}J_B$. For the simple-cubic pure $\mathbb Z_2$ gauge model, Wegner duality relates its critical coupling to that of the nearest-neighbor Ising model \citeS{S_wegner1971,S_ferrenberg2018}:
\begin{equation}
\label{eq:gauge_critical}
e^{-2K_{g,c}}=\tanh K_{I,c},
\qquad K_{I,c}=0.221654626(5),
\qquad \kappa_{\mathrm{conf}}=K_{g,c}^{-1}\simeq1.313346(3).
\end{equation}
The value $1.313346(3)$ quoted in the main text is consistent with this dual conversion at its stated precision. The perturbed critical temperature $T'_{\mathrm{conf}}=T_{\mathrm{conf}}+\delta T_{\mathrm{conf}}$ therefore obeys
\begin{equation}
\label{eq:critical_shift_equation}
\frac{J_B}{T'_{\mathrm{conf}}}
+c_4\!\left(\frac{J_A}{T'_{\mathrm{conf}}}\right)
\left(\frac{h_z}{T'_{\mathrm{conf}}}\right)^4
=\frac{J_B}{T_{\mathrm{conf}}}+O(\frac{h_z^6}{J_B^6}).
\end{equation}

Since $\delta T_{\mathrm{conf}}$ starts at fourth order, inserting it in $c_4$ or the fourth power only changes terms of eighth order or higher. Expanding the first term is therefore sufficient at the order of interest:
\begin{equation}
\label{eq:shift_expansion}
-\frac{J_B\delta T_{\mathrm{conf}}}{T_{\mathrm{conf}}^2}
+c_4(J_A/T_{\mathrm{conf}})\frac{h_z^4}{T_{\mathrm{conf}}^4}=0.
\end{equation}
Writing $r=J_A/J_B$ gives
\begin{equation}
\label{eq:alpha}
\delta T_{\mathrm{conf}}
=\alpha(r)\frac{h_z^4}{J_B^3}
+O\!\left(\frac{h_z^6}{J_B^5}\right),
\qquad
\alpha(r)=\frac{c_4(r/\kappa_{\mathrm{conf}})}{\kappa_{\mathrm{conf}}^2}.
\end{equation}

In particular, $\alpha(0)=\kappa_{\mathrm{conf}}^{-2}\simeq0.57975$. The longitudinal field raises the transition temperature at this order because it increases the effective plaquette coupling. Fig.~\ref{fig:2} of the main text shows $\alpha(r)$.

\FloatBarrier
\section{Slave-spin spectral properties and physical-electron spectroscopy}
\label{sec:s2}

In the flux-free representation, the Hamiltonian separates into free charged fermions and a cubic-lattice transverse-field Ising model,
\begin{equation}
\label{eq:decoupled_H}
H_f=-\frac t2\sum_{\langle r,r'\rangle}(f_r^\dagger f_{r'}+\mathrm{H.c.})-\mu\sum_r f_r^\dagger f_r,
\qquad
H_s=-J_s\sum_{\langle r,r'\rangle}s_r^xs_{r'}^x-h_s\sum_r s_r^z,
\quad h_s=gJ_s.
\end{equation}

Although $[H_f,H_s]=0$, the physical electron is $c_i=f_is_i^x$. Its spectrum therefore probes the slave-spin matrix elements as well as the fermion dispersion. We derive the endpoint expansions explicitly, using the Pauli normalization of the main text. Perturbative expansions for the three-dimensional transverse-field Ising model provide the broader setting for this calculation \citeS{S_weihong1994}, and the transformation used below is a block-diagonal perturbative construction \citeS{S_bravyi2011}.

The three quantities appearing in Eqs. (39)--(41) of the main text are
\begin{equation}
\label{eq:spectral_definitions}
A_0(g)=|\langle0;g|s_r^x|0;g\rangle|^2,
\qquad
A_1(g,\bm q)=N|\langle1_{\bm q};g|s_r^x|0;g\rangle|^2,
\qquad
\Delta_{g,\bm q}=E_{1,\bm q}-E_0,
\end{equation}
where $|0;g\rangle$ be the slave-spin ground state and $|1_{\bm q};g\rangle$ a normalized state in its lowest one-particle band. 
In the ordered phase, the ground state in this definition is a selected thermodynamic symmetry-broken state. This is necessary for the nonzero coherent contribution; a finite-volume parity eigenstate has vanishing $\langle s^x\rangle$ before the symmetry-breaking limit is taken. In either endpoint regime, the ground-state Lehmann representation gives, for $\tau>0$,
\begin{equation}
\label{eq:spin_lehmann}
C_s(\bm r,\tau)
=A_0+\frac1N\sum_{\bm q}A_1(g,\bm q)
e^{i\bm q\cdot\bm r-\Delta_{g,\bm q}\tau}
+\cdots
\end{equation}
The first term leaves the spin sector in its ground state, the second creates one slave-spin quasiparticle, and the remainder contains higher excitation sectors. We use the cubic harmonics
\begin{equation}
\label{eq:lattice_harmonics}
\begin{aligned}
\gamma_{\bm q}&=\cos q_x+\cos q_y+\cos q_z, \\
\xi_{\bm q}&=\cos q_x\cos q_y+\cos q_y\cos q_z+\cos q_z\cos q_x, \\
\eta_{\bm q}&=\cos2q_x+\cos2q_y+\cos2q_z.
\end{aligned}
\end{equation}
Their useful identity and Brillouin-zone averages are
\begin{equation}
\label{eq:harmonic_identity}
\gamma_{\bm q}^2=\frac32+2\xi_{\bm q}+\frac12\eta_{\bm q},
\qquad
\langle\gamma\rangle_{\mathrm{BZ}}=\langle\xi\rangle_{\mathrm{BZ}}=\langle\eta\rangle_{\mathrm{BZ}}=0.
\end{equation}

These definitions fix the numerical coefficients in the second-order results and avoid conventions in which $\gamma$ is divided by the coordination number.

\subsection{Spectral weights and excitation dispersion}

\textbf{Higgs limit $g\ll 1$}: At $g=0$, select the fully $x$-polarized state $|F\rangle=\prod_r|\rightarrow\rangle_r$. A one-defect state is $|r\rangle=s_r^z|F\rangle$. It reverses six nearest-neighbor bonds, each at energy cost $2J_s$, so
\begin{equation}
\label{eq:small_gap_bare}
|0;g=0\rangle = |F\rangle,\qquad 
|1_{\bm q};g=0\rangle = \frac{1}{\sqrt{N}} \sum_r e^{i \bm q \cdot \bm r} |r\rangle \equiv |\bm{q}\rangle ,\qquad
\Delta_{g=0,\bm{q}}=12J_s.
\end{equation}
The transverse perturbation $V=-gJ_s\sum_r s_r^z$ dresses the ground state at first order:
\begin{equation}
\label{eq:small_ground}
|0;g\rangle=|F\rangle+\frac g{12}\sum_r|r\rangle+O(g^2).
\end{equation}
Because $s_r^x$ is diagonal in this basis, $\langle r'|s_r^x|r'\rangle=1-2\delta_{r r'}$. In the normalized expectation value, the extensive vacuum normalization cancels, leaving the local reduction
\begin{equation}
\label{eq:small_A0}
\langle 0;g | s_r^x| 0;g \rangle=1-\frac{g^2}{72}+O(g^4),
\qquad
A_0=1-\frac{g^2}{36}+O(g^4).
\end{equation}

The same dressing produces the leading inelastic matrix element. At $g=0$, $s_r^x$ cannot connect $|F\rangle$ to a one-defect state. Subtracting the coherent component makes the first nonzero amplitude transparent with $\delta s_{r}^x\equiv s_r^x-\langle 0;g | s_r^x| 0;g \rangle=s_{r}^x-1 $:
\begin{equation}
\label{eq:small_connected_vertex}
\delta s_{r}^x|0;g\rangle
=-\frac{g}{6}|r\rangle+O(g^2).
\end{equation}
Since $|1_{\bm{q}};g\rangle=|\bm{q}\rangle+O(g)$ and the $O(g)$ contribution to $A_1(g,\bm{q})$ becomes of order $O(g^4)$, this yields,
\begin{equation}
\label{eq:small_A1}
\langle\bm q|\delta s_{r}^x|0;g\rangle
=-\frac{g}{6\sqrt N}e^{-i\bm q\cdot\bm r}+O(g^2),
\qquad
A_1(g,\bm q)=\frac{g^2}{36}+O(g^4).
\end{equation}
The lost coherent weight thus enters the one-particle sector at this order. The leading weight is momentum independent because the connected local operator creates a defect at one site.

To determine the excitation adiabatically connected to the one-defect
manifold at $g=0$, let $P_1=\sum_{r} \ket{r}\bra{r}$ project onto the states $|r\rangle$
and define $Q_1=1-P_1$. Since the transverse-field perturbation
$V=-gJ_s\sum_r s_r^z$ changes the number of defects by one,
$P_1VP_1=0$. The leading correction within the one-defect manifold
therefore appears at second order,
\begin{equation}
H_{\mathrm{eff}}^{(2)}
=
P_1VQ_1
\frac{1}{E_1^{(0)}-Q_1H_0Q_1}
Q_1VP_1,
\qquad
E_1^{(0)}=12J_s.
\end{equation}

The ground-state energy receives a second-order correction through the
$N$ virtual one-defect states,
\begin{equation}
\delta E_0^{(2)}
=
N\frac{(gJ_s)^2}{-12J_s}
=
-\frac{Ng^2J_s}{12}.
\end{equation}

For a defect localized at $r$, the intermediate states are the vacuum,
six adjacent two-defect states, and $N-7$ nonadjacent two-defect states.
Since the excitation energy is measured relative to the perturbed
ground state, the diagonal correction to the one-defect gap is
\begin{align}
\langle r|H_{\mathrm{eff}}^{(2)}|r\rangle
-\delta E_0^{(2)}
&=
(gJ_s)^2
\left(
\frac{1}{12J_s}
-\frac{6}{8J_s}
-\frac{N-7}{12J_s}
\right)
-\delta E_0^{(2)}
=
-\frac{g^2J_s}{12}.
\end{align}

For nearest-neighbor sites $r$ and $r'$, the defect can move through
either the vacuum or an adjacent two-defect state. The corresponding
off-diagonal matrix element is
\begin{align}
\langle r'|H_{\mathrm{eff}}^{(2)}|r\rangle
&=
(gJ_s)^2
\left(
\frac{1}{12J_s}
-\frac{1}{8J_s}
\right)
=
-\frac{g^2J_s}{24},
\qquad
r'\in\mathrm{n.n.}(r).
\end{align}
For non-neighboring sites, the vacuum and nonadjacent-pair contributions
have equal magnitudes and opposite signs and therefore cancel.

Combining the onsite correction with the nearest-neighbor hopping and
using
$\sum_{\delta}e^{i\bm q\cdot\delta}=2\gamma_{\bm q}$, where
$\gamma_{\bm q}=\cos q_x+\cos q_y+\cos q_z$, gives
\begin{equation}
\Delta_{g,\bm q}
=
J_s
\left[
12-\frac{g^2}{12}
\left(
1+\gamma_{\bm q}
\right)
\right]
+O(g^4J_s).
\end{equation}
The absence of odd powers of $g$ follows from the unitary equivalence
of $H_s(g)$ and $H_s(-g)$ under $\prod_r s_r^x$.
\begin{table}[tb]
\centering
\small
\setlength{\tabcolsep}{5pt}
\renewcommand{\arraystretch}{1.25}
\begin{tabular}{@{}lccc@{}}
\hline
\hline
Intermediate state & Excitation energy &  $E_1^{(0)}-E_m^{(0)}$ & No. of Configurations \\
\hline
Vacuum $|F\rangle$ & $0$ & $+12J_s$ &1\\
Two adjacent defect $|r,r'\rangle$ & $20J_s$ & $-8J_s$ & 6 \\
Two nonadjacent defect $|r,r'\rangle$ & $24J_s$ & $-12J_s$  &$N-7$\\
\hline
\hline
\end{tabular}
\caption{Intermediate states and their energy differences from one-defect space in the small-$g$ limit.}
\label{tab:s2}
\end{table}

\vspace{10pt}
\textbf{Orthogonal-metal limit $g\gg 1$}: At $g\gg1$, take $H_{s,0}=-h_s\sum_r s_r^z$ and use $\lambda=J_s/h_s=1/g$ as the small parameter. The vacuum is $|0\rangle=\prod_r|\uparrow\rangle_r$, and a one-defect-state $|r\rangle=s_r^x|0\rangle$ has bare cost $2h_s$. The global Ising parity $P_s=\prod_r s_r^z$ commutes with $H_s$ and reverses $s_r^x$. In the symmetry-preserving paramagnetic phase, its ground state is a parity eigenstate. Consequently,
\begin{equation}
\label{eq:large_A0}
\langle0;g|s_r^x|0;g\rangle=0,
\qquad A_0(g)=0.
\end{equation}
This vanishing is exact in the symmetric phase, rather than a truncation of the large-$g$ expansion.

Let $P_1$ project onto this one-flip manifold and define
$Q_1=1-P_1$. The perturbation is
$V=-J_s\sum_{\langle r,r'\rangle}s_r^x s_{r'}^x$.
Unlike in the small-$g$ expansion, $P_1VP_1$ is nonzero because a
bond operator incident on the flipped site moves the flip to the
other endpoint. The effective Hamiltonian through second order is
\begin{equation}
H_{\mathrm{eff}}
=
2h_sP_1
+
P_1VP_1
+
P_1VQ_1
\frac{1}{E_1^{(0)}-Q_1H_{s,0}Q_1}
Q_1VP_1
+
O\left(\frac{J_s^3}{h_s^2}\right),
\end{equation}
where $E_1^{(0)}-E_0^{(0)}=2h_s$. At first order,
\begin{equation}
\langle r'|P_1VP_1|r\rangle
=
-J_s,
\qquad
r'\in\mathrm{n.n.}(r),
\end{equation}
which gives the nearest-neighbor contribution
$-2J_s\gamma_{\bm q}$.

At second order, a bond not incident on the flipped site creates a
three-flip intermediate state. There are $3N-6$ such bonds, and the
energy denominator is $2h_s-6h_s=-4h_s$. The corresponding diagonal
matrix element is
\begin{equation}
\langle r|H_{\mathrm{eff}}^{(2)}|r\rangle
=
-\frac{(3N-6)J_s^2}{4h_s}.
\end{equation}
The ground-state correction is
\begin{equation}
\delta E_0^{(2)}
=
-\frac{3NJ_s^2}{4h_s},
\end{equation}
since each of the $3N$ bonds creates an adjacent two-flip
intermediate state with energy cost $4h_s$. The diagonal correction
to the excitation energy is therefore
\begin{align}
\langle r|H_{\mathrm{eff}}^{(2)}|r\rangle
-\delta E_0^{(2)}
&=
-\frac{(3N-6)J_s^2}{4h_s}
+\frac{3NJ_s^2}{4h_s}
=
\frac{6J_s^2}{4h_s}.
\end{align}

Two successive bond operations also move the flip to more distant
sites. An axial displacement by two lattice spacings has one
intermediate path, whereas a face-diagonal displacement has two.
Their matrix elements are
\begin{align}
\langle r+2\sigma\hat e_\alpha|
H_{\mathrm{eff}}^{(2)}
|r\rangle
&=
-\frac{J_s^2}{4h_s},
\\
\langle r+\sigma\hat e_\alpha+\sigma'\hat e_\beta|
H_{\mathrm{eff}}^{(2)}
|r\rangle
&=
-\frac{J_s^2}{2h_s},
\qquad
\alpha<\beta,
\end{align}
where $\sigma,\sigma'=\pm1$. Using the proper structure factors, these processes give
\begin{equation}
\Delta_{g,\bm q}
=
2h_s
-2J_s\gamma_{\bm q}
+
\frac{J_s^2}{4h_s}
\left(
6-8\xi_{\bm q}-2\eta_{\bm q}
\right)
+
O\left(\frac{J_s^3}{h_s^2}\right).
\end{equation}
Substituting $h_s=gJ_s$, we finally obtain
\begin{equation}
\Delta_{g,\bm q}
=
gJ_s
\left[
2-\frac{2\gamma_{\bm q}}{g}
+
\frac{6-8\xi_{\bm q}-2\eta_{\bm q}}{4g^2}
\right]
+
O\left(g^{-2}J_s\right)
.
\end{equation}

\begin{table}[t]
\centering
\begin{tabular}{@{}lccc@{}}
\hline\hline
Intermediate state
&
Excitation energy
&
$E_1^{(0)}-E_m^{(0)}$
&
No.\ of configurations
\\
\hline
One-defect state
$\lvert r'\rangle$, $r'\in\mathrm{n.n.}(r)$
&
$2h_s$
&
$0$
&
$6$
\\
Three-defect state
$\lvert r,i,j\rangle$, $\langle i,j\rangle\not\ni r$
&
$6h_s$
&
$-4h_s$
&
$3N-6$
\\
\hline\hline
\end{tabular}
\caption{Intermediate states and their energy differences from the one-defect space in the large-$g$ limit.}
\label{tab:s3}
\end{table}

We next evaluate the one-flip spectral weight $A_1(g,\bm{q})$.
To obtain the second-order coefficient in the most elegant manner, We employ a Schrieffer--Wolff transformation that block-diagonalizes the Hamiltonian with respect to the spin-flip number through second order in $1/g$. Introduce hard-core flip operators $s_r^x=b_r^\dagger+b_r$ and the two-flips-creation operator $\mathcal{C}$, two-flips-annihilation operator $\mathcal{A}$, and the hopping operator that conserves the flip number as $\mathcal{T}$.
\begin{equation}
\label{eq:flip_operators}
\mathcal C=\sum_{\langle r, r' \rangle}b_r^\dagger b_{r'}^\dagger,
\qquad \mathcal A=\mathcal C^\dagger,
\qquad \mathcal T=\sum_{\langle r, r'\rangle}(b_r^\dagger b_{r'}+b_{r'}^\dagger b_r).
\end{equation}

Then $H_s/h_s=H_{s,0}/h_s-\lambda(\mathcal C+\mathcal A+\mathcal T)$. We take an anti-Hermitian generator that removes flip-number-changing blocks through second order.
\begin{equation}
\label{eq:canonical_generator}
S=\lambda S_1+\lambda^2S_2+O(\lambda^3),
\quad S_1=\frac{\mathcal C-\mathcal A}{4},
\quad S_2=\frac{[\mathcal T,\mathcal C]+[\mathcal T,\mathcal A]}{16}.
\end{equation}

These choices cancel the flip-number-changing blocks up to $O(\lambda^2)$. We use $H_{\rm eff}=e^{-S}H_se^{S}$ and transform the measured operator $s_r^x$ with the same convention.
\begin{equation}
\label{eq:effective_spin_operator}
(s_r^x)_{\rm eff} =e^{-S} s_r^x e^S
= s_r^x+\lambda[s_r^x,S_1]
+\lambda^2\left([s_r^x,S_2]+\frac12[[s_r^x,S_1],S_1]\right)
+O(\lambda^3).
\end{equation}

Projecting onto the one-flip sector retains the local normalization and the nearest- and second-neighbor amplitudes.
\begin{equation}
\label{eq:dressed_vertex_paths}
\begin{aligned}
P_1(s_r^x)_{\rm eff}|0\rangle={}&
\left(1-\frac{3\lambda^2}{16}\right)|r\rangle
+\frac\lambda4\sum_{r' \in\mathrm{n.n.}(r)}|r'\rangle\\
&+\frac{5\lambda^2}{32}\sum_{\alpha,\sigma=\pm1}|r+2\sigma\hat e_\alpha\rangle\\
&+\frac{5\lambda^2}{16}
\sum_{\alpha<\beta}\sum_{\sigma,\sigma'=\pm1}
|r+\sigma\hat e_\alpha+\sigma'\hat e_\beta\rangle+O(\lambda^3).
\end{aligned}
\end{equation}
The hard-core algebra is important for the local term in this expression. For $|0;g\rangle = e^S|0\rangle$ and $|1_{\bm{q}};g\rangle=e^S|\bm{q}\rangle$, with $|\bm{q}\rangle = N^{-1/2} \sum_r e^{i \bm q \cdot \bm r} |r\rangle$, Fourier transform gives the transition amplitude
\begin{equation}
\label{eq:large_amplitude}
M_{\bm q} \equiv \sqrt{N} e^{i \bm q \cdot \bm r} \langle \bm{q} |(s_r^x)_{\rm eff} |0\rangle
=1+\frac\lambda2\gamma_{\bm q}
+\lambda^2\left(-\frac3{16}+\frac54\xi_{\bm q}+\frac5{16}\eta_{\bm q}\right)
+O(\lambda^3).
\end{equation}

Squaring $M_{\bm q}$ and using Eq. \eqref{eq:harmonic_identity} cancels the constant second-order term:
\begin{equation}
\label{eq:large_A1}
A_1(g,\bm q)=
|M_{\bm{q}}|^2=1+\frac{\gamma_{\bm q}}g
+\frac1{g^2}\left(3\xi_{\bm q}+\frac34\eta_{\bm q}\right)
+O(g^{-3}).
\end{equation}

\subsection{Physical electron spectral function}
Having determined the slave-spin spectrum, we insert it into the composite electron correlator. This construction is the central reason that a metallic $f$-fermion band need not imply a coherent physical-electron band \citeS{S_Nandkishore2012,S_ruegg2010}. In the decoupled flux-free density matrix, define
\begin{equation}
\label{eq:electron_greater_lesser}
G_c^>(\bm{r},t)=-i\langle c_{r}(t)c_0^\dagger(0)\rangle,
\qquad
G_c^<(\bm{r},t)=i\langle c_0^\dagger(0)c_{r}(t)\rangle.
\end{equation}

The spin operators commute with the $f$ fermions, and the two Hamiltonians commute. The thermal trace therefore factorizes:
\begin{equation}
\label{eq:factorization}
G_c^>=G_f^>C_s^>,
\qquad G_c^<=G_f^<C_s^<,
\end{equation}
where $C_s^>(\bm{r},t)=\langle s_{r}^x(t)s_0^x(0)\rangle$ and $C_s^<(\bm{r},t)=\langle s_0^x(0)s_{r}^x(t)\rangle$. This factorization is exact in the decoupled sector; truncating its spin Lehmann representation is the approximation.

For $\epsilon_{\bm q}=-t\gamma_{\bm q}-\mu$ and defining  $\int_{\bm q}=\int_{\mathrm{BZ}}d^3q/(2\pi)^3$, the free-fermion factors are
\begin{equation}
\label{eq:free_greater_lesser}
\begin{aligned}
G_f^>(\bm{r},t)&=-i\int_{\bm q}[1-n_F(\epsilon_{\bm q})]
e^{i\bm q\cdot\bm{r}-i\epsilon_{\bm q}t},\\
G_f^<(\bm{r},t)&= i\int_{\bm q}n_F(\epsilon_{\bm q})
e^{i\bm q\cdot\bm{r}-i\epsilon_{\bm q}t},
\end{aligned}
\end{equation}

We retain these Fermi functions throughout. Under $T\ll\Delta_g=\min_{\bm p}\Delta_{g,\bm p}$, thermal slave-spin occupation is exponentially small, so the spin factors can be evaluated with ground-state matrix elements:
\begin{equation}
\label{eq:spin_greater_lesser}
\begin{aligned}
C_s^>(\bm{r},t)&\simeq A_0+\int_{\bm p}A_1(g,\bm p)
e^{i\bm p\cdot\bm{r}-i\Delta_{g,\bm p}t},\\
C_s^<(\bm{r},t)&\simeq A_0+\int_{\bm p}A_1(g,\bm p)
e^{i\bm p\cdot\bm{r}+i\Delta_{g,\bm p}t}.
\end{aligned}
\end{equation}

In writing the second line we used inversion symmetry, $A_1(\bm p)=A_1(-\bm p)$ and $\Delta_{\bm p}=\Delta_{-\bm p}$. Multiplication in real space gives a momentum convolution with slave-spin momentum $\bm p=\bm k-\bm q$. Fourier transforming $G_c^>-G_c^<$ gives
\begin{equation}
\label{eq:electron_convolution}
\begin{aligned}
A_c(\bm k,\omega)={}&A_0(g)\delta(\omega-\epsilon_{\bm k})\\
&+\int_{\bm q}A_1(g,\bm k-\bm q)[1-n_F(\epsilon_{\bm q})]
\delta(\omega-\epsilon_{\bm q}-\Delta_{g,\bm k-\bm q})\\
&+\int_{\bm q}A_1(g,\bm k-\bm q)n_F(\epsilon_{\bm q})
\delta(\omega-\epsilon_{\bm q}+\Delta_{g,\bm k-\bm q}).
\end{aligned}
\end{equation}

The normalization is $A_c=-\pi^{-1}\operatorname{Im}G_c^R$. Equation \eqref{eq:electron_convolution} reproduces Eq. (45) of the main text. The upper sideband adds an $f$ fermion and a spin excitation; the lower sideband removes an $f$ fermion while creating a spin excitation. Higher spin sectors and thermally occupied slave-spin states provide corrections to the stated one-particle, low-temperature approximation.

As $T\ll t \sim \epsilon_{\bm q} $, $1-n_F(\epsilon)\to\Theta(\epsilon)$ and $n_F(\epsilon)\to\Theta(-\epsilon)$. Keeping $n_F$ before this limit is essential whenever the fermionic band is thermally resolved. Integrating over frequency provides a second check:
\begin{equation}
\label{eq:electron_sum_rule}
\int_{-\infty}^{\infty}d\omega\,A_c(\bm k,\omega)\Big|_{T\ll t} \approx
 A_0+\int_{\bm q}A_1(g,\bm k-\bm q)
=\begin{cases}
1+O(g^4),&g\ll1,\\
1+O(g^{-3}),&g\gg1.
\end{cases}
\end{equation}
to the perturbative and thermal accuracy of the retained spin sectors. The cancellation of the two Fermi factors in this integrated sum confirms that they implement the internal particle and hole occupations consistently.

\section{Self-energy from temporal star fluctuations}
\label{sec:s3}
To distinguish the original orthogonal-metal model from a generic
metallic band, we include a gauge-independent hopping $t_0$:
\begin{align}
H_f
={}&
-\frac{1}{2}
\sum_{\langle r,r'\rangle}
\left(t_0+t\sigma^z_{rr'}\right)
\left(c_r^\dagger c_{r'}+\mathrm{H.c.}\right)
-\mu\sum_r n_r,
\\
H_s
={}&
-J_s\sum_{\langle r,r'\rangle}\sigma^z_{rr'}
-h\sum_r P_rA_r,
\end{align}
where
\begin{equation}
P_r=(-1)^{n_r},
\qquad
A_r=
\prod_{r'\in\mathrm{n.n.}(r)}
\sigma^x_{rr'}.
\end{equation}
The Hamiltonian of the main text corresponds to $t_0=0$. We consider
the hierarchy
\begin{equation}
E_c=t+t_0\gg T\gg J_s,h,
\end{equation}
so that temporal star fluctuations may be treated perturbatively.

We condition on an initial link configuration $u$ and consider a
closed temporal star history. The leading local process contains two
star insertions at the same site:
\begin{equation}
u
\xrightarrow{\,A_r(q_2)\,}
u^{(r)}
\xrightarrow{\,A_r(q_1)\,}
u,
\qquad
\tau=q_1-q_2>0.
\end{equation}
Here, $u^{(r)}$ is obtained by reversing the six links incident on
$r$. A history with a single local star insertion does not close under
the thermal trace. The leading self-energy correction is therefore of
order $h^2$.

Let
\begin{equation}
H_0=H_f[u],
\qquad
H_1=P_rH_f[u^{(r)}]P_r.
\end{equation}
The parity transformation reverses the sign of the fermion operator
at site $r$, giving
\begin{equation}
H_1=H_0+t_0W_r,
\qquad
W_r=
\sum_{j\in\mathrm{n.n.}(r)}
\left(
c_r^\dagger c_j+c_j^\dagger c_r
\right).
\end{equation}
Thus, during the interval $\tau$, the fermions experience a local
quench proportional to $t_0$. The corresponding many-body overlap is
\begin{equation}
D_x(\tau)
=
\frac{
\operatorname{Tr}_f
\left[
e^{-(\beta-\tau)H_0}e^{-\tau H_1}
\right]
}{
\operatorname{Tr}_f e^{-\beta H_0}
},
\qquad
x=\frac{t}{t+t_0}.
\end{equation}

For a uniform metallic background, the quench is restricted to the
two-dimensional single-particle subspace spanned by
\begin{equation}
|r\rangle,
\qquad
|s_r\rangle
=
\frac{1}{\sqrt{6}}
\sum_{j\in\mathrm{n.n.}(r)}
|j\rangle.
\end{equation}
Evaluating the associated rank-two scattering determinant gives
\begin{equation}
d_x(z;\mu)
=
(2x-1)^2
+
4x(1-x)(z+\mu)G_{\mathrm{loc}}(z;\mu),
\end{equation}
where
\begin{equation}
G_{\mathrm{loc}}(z;\mu)
=
\frac{1}{N}
\sum_{\bm{k}}
\frac{1}{
z+\mu+E_c\gamma_{\bm{k}}
},
\qquad
\gamma_{\bm{k}}
=
\cos k_x+\cos k_y+\cos k_z.
\end{equation}

At half filling and for a generic mixture $0<x<1$, the scattering
phase obtained from $d_x(z;0)$ gives a positive quench threshold,
\begin{equation}
\Delta_\star(x)>0.
\end{equation}
The corresponding many-body overlap has the long-time asymptotic form
\begin{equation}
D_x(\tau)
\sim
\mathcal{A}_x
e^{-\Delta_\star(x)\tau}
(E_c\tau)^{-\alpha_{\mathrm{AO}}(x)},
\qquad
E_c^{-1}\ll\tau\ll\beta,
\end{equation}
where $\alpha_{\mathrm{AO}}(x)$ is the Anderson-orthogonality exponent.
The positive threshold exponentially suppresses long star intervals,
so that the time moments of $D_x(\tau)$ are finite.

For a normalized Bloch state for local basis $\{|r\rangle,|s_r\rangle\}$, the projected finite temperature host Green function is
\begin{eqnarray}
G_c(\tau) = \frac{1}{N} \sum_{\bm k} \bm{u}_{\bm{k}} \bm{u}_{\bm{k}}^{\dagger} g_{\bm k} (\tau),\quad g_k(\tau) = \begin{cases} -[1-n_F(\epsilon_{\bm k})] e^{-\epsilon_{\bm k} \tau},\ &0<\tau<\beta, \\ n_F(\epsilon_{\bm k})e^{-\epsilon_{\bm k} \tau},\ &-\beta<\tau<0. \end{cases},\quad \bm{u_k}=\begin{pmatrix} 1 \\ \sqrt{2/3}\gamma_{\bm k} \end{pmatrix}.
\end{eqnarray}
For the quadratic reference Hamiltonian $H_0$, we define
\begin{eqnarray}
G_{ab}^{(0)} (\tau) = - \big \langle T_{\tau} c_a(\tau)c_b^{\dagger}(0) \big \rangle_0.
\end{eqnarray}
Applying the parity transformation to the full fixed-$\tau$ external line gives
\begin{eqnarray}
\label{eq:full fixed external line}
G_{\bm k k }^{(\tau)}(s)-2\chi_{(0,\tau)}(s) G_{r \bm{k}}^{(\tau)}(s) 
= \big [G_{\bm{k k}}^{(0)}(s)-2\chi_{(0,\tau)}(s) G_{r \bm{k}}^{(0)}(s)\big]
+\big[\delta G_{\bm{k k}}^{(\tau)}(s)-2\chi_{(0,\tau)}(s) \delta G_{r \bm{k}}^{(\tau)}(s)\big].
\end{eqnarray}
where $\chi_{(0,\tau)}(s)=1$ for $0<s<\tau$ and vanishes otherwise.

The first bracket of Eq. \eqref{eq:full fixed external line} gives the parity-transformed bare external line. Using
\begin{equation}
P_rc_{\bm{k}}P_r
=
c_{\bm{k}}
-
\frac{2e^{-i\bm{k}\cdot\bm{R}_r}}{\sqrt{N}}c_r,
\end{equation}
the parity contribution to the leading self-energy at the Fermi
surface becomes
\begin{align}
\Sigma_P(\bm{k}_F,i\omega_n)
&=
-2ih^2
\int_0^\beta d\tau\,
D_x(\tau)\sin(\omega_n\tau)
\nonumber\\
&=
-2ih^2\omega_n
\int_0^\beta d\tau\,
\tau D_x(\tau)
+
O(h^2\omega_n^3,h^4).
\end{align}

The second bracket of Eq. \eqref{eq:full fixed external line} contains the quench-induced open-line correction. On the interval $0<s<\tau$, the bare external source $R_0(s)=\bm{u}_{\bm{k}_F} g_{\bm k_F} (s)$ under repeated scattering is summed by
\begin{equation}
Y_\tau(s) =R_0(s) + \int_0^\tau ds' G_c(s-s')V_x Y_\tau(s'),\quad Z_\tau(s)=V_xY_\tau(s),
\end{equation}
where $t_0 W_r\doteq V_x=\sqrt{6}t_0\sigma_x$ in the local basis $\{|r\rangle,|s_r\rangle\}$. We define two finite-interval transforms
\begin{eqnarray}
\mathcal{A}_\tau(z) = \int_0^\tau ds\, e^{zs} \bm{u}_{\bm{k}_F}^{\dagger} Z_\tau(s),\quad \mathcal{B}_\tau(z)= \int_0^\tau ds\, e^{zs}\big[Y_\tau(s)-R_0(s)\big]_r.
\end{eqnarray}
The open-line contribution to the leading self-energy at the Fermi
surface becomes
\begin{align}
\Sigma_{open}(\bm{k}_F,i\omega_n)
&=
ih^2\omega_n
\int_0^\beta d\tau\,
D_x(\tau)\big[e^{i\omega_n(\beta-\tau)}-1\big] \big[g_{\bm{k}_F}(i\omega_n) \mathcal{A}_\tau(i\omega_n) - \mathcal{B}_\tau(i\omega_n)\big]
\nonumber\\
&=
ih^2\omega_n\ \mathrm{Re}
\int_0^\beta d\tau\,
D_x(\tau) [\tau \mathcal{A}_\tau^{(0)} -2\mathcal{A}_\tau^{(1)} + 4 \mathcal{B}_\tau^{(0)}]
+
O(h^2\omega_n^2,h^4).
\end{align}
where
\begin{eqnarray}
\mathcal{A}_\tau^{(m)} = \int_0^\tau ds\, s^m \bm{u}_{\bm{k}_F}^{\dagger}Z_\tau(s),\quad \mathcal{B}_{\tau}^{(0)} = \int_0^\tau ds\,[Y_\tau(s)-R_0(s)]_r. 
\end{eqnarray}
For fixed $0<x<1$, these parity-induced  and open-line terms give
\begin{equation}
\Sigma(\bm{k}_F,i\omega)
-
\Sigma(\bm{k}_F,0)
\propto
\frac{h^2}{E_c^2}\,i\omega.
\end{equation}

At the lowest fermionic Matsubara frequency, provided
$\pi T\ll\Delta_\star(x)$,
\begin{equation}
-\operatorname{Im}
\left[
\Sigma(\bm{k}_F,i\pi T)
-
\Sigma(\bm{k}_F,0)
\right]
\propto
\frac{h^2T}{E_c^2},
\qquad
0<x<1.
\end{equation}

The two endpoints must be treated separately because the positive
threshold controlling the generic mixed-hopping regime vanishes
there. At $x=0$, corresponding to $t=0$, the two Hamiltonians are
unitarily related:
\begin{equation}
H_1=P_rH_0P_r.
\end{equation}
The leading self-energy is consequently proportional to the local
Green function of the direct-hopping band,
\begin{equation}
G_{t_0}^{\mathrm{loc}}(z)
\equiv
\left.
G_{\mathrm{loc}}(z;\mu)
\right|_{t=0},
\qquad
\Sigma(z)
=
4h^2G_{t_0}^{\mathrm{loc}}(z)
+
O(h^4).
\end{equation}
For a metallic density of states, the low-frequency Green function
contains a nonanalytic $\operatorname{sgn}\omega$ contribution.
Consequently, the lowest-Matsubara self-energy is temperature
independent to leading order:
\begin{equation}
-\operatorname{Im}\Sigma(i\pi T)
\propto
h^2,
\qquad
x=0.
\end{equation}

At $x=1$, corresponding to $t_0=0$ and hence to the original model,
the local quench disappears:
\begin{equation}
H_1=H_0,
\qquad
D_1(\tau)=1.
\end{equation}
The parity contribution is therefore
\begin{align}
\Sigma_P(\bm{k}_F,i\omega_n)
&=
-2ih^2
\int_0^\beta d\tau\,
\sin(\omega_n\tau)
\nonumber\\
&=
\frac{4h^2}{i\omega_n}.
\end{align}
Thus, the leading self-energy is
\begin{equation}
\Sigma(\bm{k}_F,i\omega_n)
=
\frac{4h^2}{i\omega_n}
+
O(h^4).
\end{equation}
At the lowest Matsubara frequency,
\begin{equation}
-\operatorname{Im}\Sigma(\bm{k}_F,i\pi T)
=
\frac{4h^2}{\pi T}
+
O(h^4).
\end{equation}
The results at three different regime are summarized in Table~\ref{tab:s4}.

\FloatBarrier
\begin{table}[h]
\centering
\setlength{\tabcolsep}{9pt}
\renewcommand{\arraystretch}{1.1}
\begin{tabular}{@{}ccc@{}}
\hline\hline
Regime
&
Quench structure
&
Lowest-Matsubara response
\\
\hline
$x=0$ $(t=0)$
&
$H_1=P_rH_0P_r$
&
$-\operatorname{Im}\Sigma(i\pi T)\propto h^2$
\\
$0<x<1$
&
$\Delta_\star(x)>0$
&
$-\operatorname{Im}\delta\Sigma(i\pi T)
\propto h^2T/E_c^2$
\\
$x=1$ $(t_0=0)$
&
$H_1=H_0$
&
$-\operatorname{Im}\Sigma(i\pi T)
=4h^2/(\pi T)$
\\
\hline\hline
\end{tabular}
\caption{Lowest-Matsubara self-energy generated by temporal star fluctuations
at the direct-hopping endpoint, in the generic mixed-hopping regime,
and in the original orthogonal-metal model. Here,
$\delta\Sigma(i\omega)=\Sigma(i\omega)-\Sigma(0)$.
}
\label{tab:temporal-star-scaling}
\label{tab:s4}
\end{table}

\FloatBarrier
\begingroup
\makeatletter
\let\NAT@bibsetnum\NATx@bibsetnum
\let\@bibsetup\NATx@bibsetnum
\setbiblabelwidth{99}
\let\savedprebibdata\pre@bibdata
\let\pre@bibdata\@empty
\bibliographystyleS{apsrev4-2}
\bibliographyS{refs_supp}
\let\pre@bibdata\savedprebibdata
\makeatother
\endgroup

\end{document}